\documentclass{aa}
\usepackage{graphicx}
\usepackage{bm,natbib,url,wasysym}
\usepackage[varg]{txfonts}
\usepackage{xcolor}
\usepackage{amsmath}
\usepackage[colorlinks,allcolors=blue]{hyperref}
\bibpunct{(}{)}{;}{a}{}{,} % to follow the A&A style
\graphicspath{{./figs/}{./png/}}
\newcommand{\brac}[1]{\langle #1 \rangle}
\newcommand{\EQ}{\begin{equation}}
\newcommand{\EN}{\end{equation}}
\newcommand{\EQA}{\begin{eqnarray}}
\newcommand{\ENA}{\end{eqnarray}}

\newcommand{\Eq}[1]{Equation~(\ref{#1})}

\newcommand{\Eqss}[2]{Equations~(\ref{#1})--(\ref{#2})}

\newcommand{\Sec}[1]{Section~\ref{#1}}

\newcommand{\Fig}[1]{Fig.~\ref{#1}}

\newcommand{\Figs}[2]{Figs.~\ref{#1} and \ref{#2}}

\newcommand{\Tab}[1]{Table~\ref{#1}}

\newcommand{\bbra}[1]{\left\langle #1\right\rangle}
\newcommand{\fluc}[1]{#1^\prime}

\newcommand{\mean}[1]{\overline#1}
\newcommand{\meanrho}{\overline{\rho}}

{}
{}
{}

\newcommand{\meanuu}{\overline{\mbox{\boldmath $u$}}{}}{}
{}
{}
{}
{}
{}
{}
{}
{}
\newcommand{\meanBB}{\overline{\mbox{\boldmath $B$}}{}}{}
{}
{}
{}
{}
{}
{}
{}
{}
\newcommand{\meanUU}{\overline{\bm{U}}}

{}
{}
{}

\newcommand{\meanB}{\overline{B}}

\newcommand{\meanU}{\overline{U}}

\newcommand{\flucbb}{\mbox{\boldmath $b^\prime$}{}}{}
\newcommand{\flucuu}{\mbox{\boldmath $u^\prime$}{}}{}
{}
{}

{}

{}
{}

\newcommand{\alphaK}{\alpha_{\rm K}}
\newcommand{\alphaM}{\alpha_{\rm M}}
\newcommand{\alphaKM}{\alpha_{\rm KM}}
\newcommand{\alpharr}{\alpha_{rr}}
\newcommand{\alphatt}{\alpha_{\theta\theta}}
\newcommand{\alphapp}{\alpha_{\phi\phi}}

\newcommand{\Ot}{\tilde{\Omega}}

\newcommand{\uu}{\mbox{\boldmath $u$} {}}

\newcommand{\BB}{\mbox{\boldmath $B$} {}}

\newcommand{\jj}{\mbox{\boldmath $j$} {}}

\newcommand{\AAA}{\mbox{\boldmath $A$} {}}

\newcommand{\nab}{\mbox{\boldmath $\nabla$} {}}
\newcommand{\OO}{\bm{\Omega}}
\newcommand{\oo}{\mbox{\boldmath $\omega$} {}}

\newcommand{\ddelta}{\mbox{\boldmath $\delta$} {}}
\newcommand{\ggamma}{\mbox{\boldmath $\gamma$} {}}
\newcommand{\aalpha}{\mbox{\boldmath $\alpha$} {}}
\newcommand{\bbeta}{\mbox{\boldmath $\beta$} {}}
\newcommand{\kkappa}{\mbox{\boldmath $\kappa$} {}}

\newcommand{\EMF}{\mbox{\boldmath ${\cal E}$} {}}

\def\Ta{\mbox{\rm Ta}}
\def\Ra{\mbox{\rm Ra}}

\def\Co{\mbox{\rm Co}}

\def\PrSGS{\mbox{\rm Pr}_{\rm SGS}}

\def\Pm{\mbox{\rm Pr}_{\rm M}}
\def\Rm{\mbox{\rm Re}_{\rm M}}

\def\Rey{\mbox{\rm Re}}

\def\Co{\mbox{\rm Co}}

\def\kf{k_{\rm f}}

\def\urms{u_{\rm rms}}
\def\urmsp{u^\prime_{\rm rms}}

\def\etatz{\eta_{\rm t0}}

\def\half{{\textstyle{1\over2}}}

\newcommand{\s}{\,{\rm s}}

\newcommand{\m}{\,{\rm m}}

\newcommand{\kg}{\,{\rm kg}}

\newcommand{\chiSm}{\chi_{\rm m}^{\rm SGS}}

\begin{document}

\titlerunning{Rotational dependence of turbulent transport coefficients}
\authorrunning{Warnecke \& K\"apyl\"a}

\title{Rotational dependence of turbulent transport coefficients\\in 
global convective dynamo simulations of solar-like stars}
\author{J. Warnecke\inst{1} 
\and M. J.\  K\"apyl\"a\inst{1,2}}
\institute{Max-Planck-Institut für Sonnensystemforschung,
  Justus-von-Liebig-Weg 3, D-37077 G\"ottingen, Germany\\
\email{warnecke@mps.mpg.de}\label{inst1} 
\and ReSoLVE Centre of Excellence, Department of Computer Science, Aalto University, PO Box 15400, FI-00\ 076 Aalto,
Finland \label{inst2}
}

%\date{\today,~ $ $Revision: 1.70 $ $}
\date{Received 15 October 2019 / Accepted 25 February 2020}
\abstract{
For moderate and slow rotation, the magnetic activity of solar-like stars 
is observed to strongly depend on rotation,
while for rapid rotation, only a very weak or no dependency is 
detected. 
These observations do not yet have a solid explanation in terms of dynamo theory.
}{
We aim to find such an explanation by numerically 
investigating the rotational dependency of dynamo 
drivers in solar-like stars, that is, stars that have a convective envelope of similar thickness to that of the Sun.
}{
We ran semi-global convection simulations of stars with rotation
  rates from 0 to 30   times the solar value, corresponding to
  Coriolis numbers, $\Co$, of 0 to 110. We measured the turbulent
  transport coefficients describing the magnetic field evolution with
  the help of the  test-field method,
  and compared with the dynamo effect arising from the differential
  rotation that is self-consistently generated in the models.
}{The trace of the $\aalpha$ tensor increases for moderate 
rotation rates with Co$^{0.5}$ and levels off for rapid rotation. 
This behavior is in agreement with
  the kinetic $\alpha$ based on the kinetic helicity, if one 
  takes into account the decrease of the convective scale
 with increasing rotation.
  The $\aalpha$ tensor
  becomes highly anisotropic for $\Co\gtrsim 1$. Furthermore, $\alpharr$ dominates
  for moderate rotation (1<$\Co$<10), and $\alphapp$ for rapid rotation
  ($\Co \gtrsim 10$). 
    The effective meridional flow, taking into account the 
    turbulent pumping effects,
  is markedly different from the actual meridional circulation profile.
  Hence, the turbulent pumping effect is
  dominating the  meridional 
  transport of the magnetic field.
  Taking all dynamo effects
  into account, we find three distinct regimes. For slow rotation, the
  $\alpha$ and R\"adler effects are dominating in the presence of anti-solar
  differential rotation. 
  For moderate rotation, $\alpha$ and $\Omega$ effects are 
  dominant, indicative of   $\alpha\Omega$ or $\alpha^2\Omega$ dynamos in operation,
  producing  equatorward-migrating dynamo waves with a qualitatively solar-like rotation profile.
  For rapid rotation, an $\alpha^2$ 
mechanism with an influence from the R\"adler 
effect appears to be the most probable driver of the dynamo.
}{Our study reveals the presence of a large variety of dynamo effects
  beyond the classical $\alpha\Omega$
   mechanism, which need to be investigated
  further to fully understand the dynamos of solar-like stars.
  The highly anisotropic $\aalpha$ tensor might be the primary
  reason for the change of axisymmetric to non-axisymmetric dynamo
  solutions in the moderate rotation regime.
 }


\maketitle

\section{Introduction}

The magnetic activity of stars shows a strong dependency on rotation, which is most
pronounced  in their coronal X-ray flux
\citep[e.g.,][]{PMMSBV03,VGJDPMFBC14,RSP14,WD16} and their
chromospheric Ca II H\&K emission
\citep[e.g.,][]{Noyes+al:1984b,BST1998,BMJRCMPW17,OLKPG17}.
Higher rotation, usually measured using the Coriolis number, Co, describing the 
rotational influence on convection leads to stronger
coronal and chromospheric emission.
For rapidly rotating stars with Co $\gtrsim 10$ \citep[e.g.,][]{WD16},
the emission becomes independent of rotation,
which is often referred to as the ``saturated'' regime. 
This terminology is somewhat misleading, as it may be confused with the nonlinear
saturation of the dynamo; all the dynamos in these stars are indeed expected to be
saturated dynamos.
The coronal and chromospheric emission can be linked to the surface
magnetic field
\citep[e.g.,][]{PFALJ03,VGJDPMFBC14}, and therefore to the underlying
dynamo process.
Therefore, it is very important to study the dependency of the dynamo process
on rotation.

The most important dynamo  effect 
in astrophysical systems is the
$\alpha$ effect \citep{SKR66}, 
which describes the ability of small-scale velocity with a
twist, for example due to rotation in a stellar convection zone, 
to amplify the magnetic field.
Using mean-field theory with the second-order correlation
approximation (SOCA) and assuming isotropic homogeneous turbulence,
\cite{SKR66} find that $\alpha$ increases linearly with rotation in the slow
rotation
regime because the
$\alpha$ effect is directly related to the kinetic helicity under
these assumptions.
An extension of the theory to higher rotation predicts that the $\alpha$ will level off 
in this regime \citep[e.g.,][]{RK93}.
Differential rotation, the other important ingredient of a
stellar dynamo process, is predicted to depend only weakly on rotation
using the models by \cite{KR99}.
In these models, turbulent effects generating the differential
rotation are parametrized and obtained from mean-field theory; hence the common reference to ``mean-field models''.
The weak rotational dependence of differential rotation 
is confirmed by observational studies of the
surface latitudinal differential rotation \citep[e.g.,][]{RRB13,LJHKH16}.

Global convective dynamo simulations 
have been able to identify three
distinctive regimes in terms of rotation. At moderate rotation, where the
Coriolis number is between three and ten, these simulations 
produce cyclic dynamo waves propagating towards the equator
\citep[e.g.,][]{GCS10,KMB12,ABMT15,SBCBN17,W18}, a basic feature
of the solar magnetic field evolution.
Recently some of these simulations have even reproduced dynamo
solutions with multiple modes with shorter and longer periods than the
dominant cycle \citep{BSCC16,KKOBWKP16,KKOWB17}.
Most of these cyclic dynamo solutions can be
explained by an $\alpha\Omega$ dynamo wave following the
Parker-Yoshimura rule
\citep{P55,Yos75} as shown in the studies by \cite{WKKB14,WRTKKB17}, \cite{KKOBWKP16,KKOWB17}, and
\cite{W18}.
In a classical $\alpha\Omega$ dynamo, the poloidal magnetic field is
generated by the $\alpha$ effect from the toroidal magnetic field,
which is produced from the shear of the differential rotation
($\Omega$ effect).
To excite an equatorward-migrating waves, such as the one seen in the Sun, the product of
$\alpha$ and the radial shear must be negative(positive) in
northern(southern) hemisphere.
In these simulations, the sign of $\alpha$ is unfavorable
for the correct migration direction of the dynamo wave in the bulk of the
convection zone. Instead, most of the simulations produce a local minimum of
negative shear, which results in the correct migration direction. However, such
a feature is not seen in solar observations, 
although negative shear is present in the very topmost layer of the
convection zone,  called the near-surface shear layer
\citep{Thompson96,BSG14}.
Only in the work of \cite{DWBF15}, has equatorward migration been seen
in thick convection zones. This was found to result 
from the reversed sign of kinetic helicity, hence the $\alpha$, in the 
bulk of the convection zone.

At Coriolis numbers around unity and below,
the differential rotation profile develops 
fast poles and a slow equator, which is opposite to the Sun, with its
fast equator and slow poles; hence the name, anti-solar differential
rotation profile. 
In this regime, most of the simulations produce
irregular in-time dynamo solutions \citep{KKKBOP2015,W18}. However,
\cite{VWKKOCLB17,VKWKR19} 
 discovered a cyclic solution in this regime.
None of these
dynamo solutions 
can be explained by a pure $\alpha\Omega$ dynamo
as in the moderate rotation case. The study of \cite{VKWKR19} revealed
that the $\alpha$ effect generating the toroidal magnetic field is
comparable to or even larger than the $\Omega$ effect of differential
rotation.
In the Coriolis number range between the regime described above, we  often
find a mixture of both dynamo types \citep[e.g.,][]{VWKKOCLB17,W18}.

For large Coriolis numbers, 
mean-field dynamo models predict dynamo solutions with non-axisymmetric large-scale magnetic
field which is often associated with a strong anisotropy of the $\aalpha$
tensor \citep[e.g.,][]{RWBMT90,ER07,P17}.
However, non-axisymmetric dynamo solutions have also been obtained with an
isotropic $\aalpha$ \citep[e.g.,][]{MB95,MBBT95,tuominen2002starspot}. 
Observational studies also indicate strong nonaxisymmetric 
surface field \citep[e.g.,][]{MDPDFJ10} or photometric spot
distribution \citep{LJHKH16} for stars with a high Coriolis number.
Global convective dynamo simulations confirm nonaxisymmetric dynamo
solutions for moderately and rapidly rotating stars
\citep{KMCWB13,Coletal14,VWKKOCLB17}. 
The dynamo drivers have not yet been systematically
measured as a function of rotation for these simulations.
In particular, we are interested 
in whether or not the $\aalpha$ tensor becomes
anisotropic in these simulations.
These are the main purposes of the present paper.

We use the
test-field method \citep{SRSRC05,SRSRC07} to determine the 
turbulent transport
coefficients.
This method has been shown
to give a good description of the dynamo processes in global dynamo
simulations at moderate Reynolds numbers
\citep{Schr11,SPD11,SPD12,WRTKKB17,W18, VKWKR19}.
As the current test-field method only works for cases where the
large-scale magnetic field is axisymmetric, we restrict our setup
to azimuthal wedges of one-quarter of a sphere. Hence, large-scale
nonaxisymmetric modes are suppressed and we can study the rotational
dependency of the turbulent transport coefficients independently of
other parameters.
The dataset containing the turbulent transport coefficients as well as
the mean flows for all runs can be found at
\url{http://doi.org/10.5281/zenodo.3629665}.

\section{Model and setup}
\label{sec:model}

We model the stellar 
convection zone in a spherical wedge
($r,\theta,\phi$), with the same depth as the Sun, ($r=0.7\, R$ to $r=R$),
where $R$ is the stellar radius. We restrict our domain to a quarter of
a sphere ($0\le\phi\le\pi/2$) without poles
($\theta_0\le\theta\le\pi-\theta_0$, 
where $\theta_0=15^{\circ}$)
for numerical reasons.
We solve 
equations of magnetohydrodynamics in a fully compressible regime,
including the induction equation for the magnetic field $\BB$ in terms
of the vector potential $\AAA$, which ensures the solenoidality of
$\BB=\nab\times\AAA$, the momentum equation in terms of the velocity
$\uu$, the 
continuity equation for the density $\rho$,
and the energy
equation in terms of the specific entropy $s$ together with an
equation of state for an ideal gas with temperature $T$.
Rotation is included via the Coriolis force $\OO_0\times\uu$, where
$\OO_0=\Omega_0(\cos\theta,-\sin\theta,0)$ is the rotation vector with
the bulk rotation $\Omega_0$,
 and gravity via Keplerian
acceleration. The plasma is heated by 
a constant heat flux at the bottom
of the convection zone and is cooled at the top by invoking a black-body
boundary condition. We use a periodic boundary condition
in the azimuthal direction for all quantities, a stress-free 
condition for the velocity on all other boundaries,
and a perfect conductor condition for the magnetic
field at the bottom radial and the latitudinal boundaries. At the top
radial boundary, the magnetic field is radial.
Further details of the model setup are described in \cite{KMCWB13} and
will not be repeated here.

Our model is characterized by non-dimensional input
parameters: the normalized 
rotation rate and the resulting Taylor number,
\begin{equation}
\Ot=\Omega_0/\Omega_\odot, \quad \Ta=[2\Omega_0 (0.3R)^2/\nu]^2,
\end{equation}
where $\Omega_\odot=2.7\times10^{-6} \s^{-1}$ is the rotation rate of
the Sun, and $\nu$ is the constant
kinematic viscosity,  and the sub-grid-scale thermal
and magnetic Prandtl numbers
are\begin{equation}
\PrSGS={\nu\over\chiSm},\quad \Pm={\nu\over\eta}, 
\end{equation}
where $\chiSm$ is the sub-grid-scale thermal diffusivity in the middle
of the convection zone, and $\eta$ is the constant magnetic diffusivity.
Additionally, we define  
a turbulent Rayleigh number calculated from a
hydrostatic one-dimensional model
\begin{equation}
\Ra\!=\!\frac{GM(0.3R)^4}{\nu \chiSm R^2}
  \bigg(-\frac{1}{c_{\rm P}}\frac{{\rm d}s_{\rm hs}}{{\rm d}r}
  \bigg)_{(r=0.85R)},
\label{equ:Ra}
\end{equation}
where $s_{\rm hs}$ is the specific entropy in the hydrostatic model,
$G$ is the gravitational acceleration, $M$ is the mass of the star, and
$c_{\rm P}$ is the specific heat capacity at constant pressure.

We use the fluid and magnetic Reynolds numbers together with the
Coriolis number
\begin{equation}
\Rey=\frac{\urms}{\nu \kf},\quad \Rm=\frac{\urms}{\eta \kf},\quad
\Co={2\Omega_0\over\urms\kf},
\end{equation}
to characterize our simulations.
Here, $\kf=2\pi/0.3R\approx21/R$ is an estimate of the wavenumber of
the largest eddies in the convection 
zone and $\urms=\sqrt{(3/2)\brac{u_r^2+u_\theta^2}_{r\theta\phi t}}$ is
the rms velocity and the subscripts indicate averaging over $r$,
$\theta$, $\phi$ and a time interval covering the saturated state.
These non-dimensional inputs and characteristic parameters are given
in \Tab{runs}.

For our analysis we divide each field into a mean (axisymmetric) part and
a fluctuating part, the mean being denoted 
with an overbar and the
fluctuations with a prime, for example, $\BB=\meanBB+\fluc{\BB}$ and
$\uu=\meanuu+\fluc{\uu}$. We note that this axisymmetric mean follows
the Reynolds rules.
As we are using the wedge approximation 
in the azimuthal direction, the
large-scale nonaxisymmetric modes with azimuthal degrees of 1, 2, and 3
are suppressed. 
Hence, the adopted azimuthal mean can be reliably used to
compute the mean fields, which accurately describes the large-scale magnetic
field evolution.
With this azimuthal mean, we define a $r$- and $\theta$-dependent turbulent velocity
as $\urmsp(r,\theta)={\left\langle\,\overline{{\bm
        u}^{\prime\,2}}\,\right\rangle_t}{}^{\!\!1/2}$ and the
corresponding turnover time of the convection $\tau_{\rm
  tur}=H_p\alpha_{\rm MLT}/\urmsp$, where
$H_p=-(\partial \ln\mean{p}/\partial r)^{-1}$ is the pressure scale
height and $\alpha_{\rm MLT}=5/3$ is the mixing length parameter.

We define the total kinetic energy as
\begin{equation}
E_{\rm kin}^{\rm tot}=\half\bbra{\rho\uu^2}_V,
\label{eq:ene1}
\end{equation}
which can be decomposed into energies of the fluctuating velocities, the
differential rotation, and the meridional circulation:
\begin{eqnarray}
&&E_{\rm kin}^{\rm flu} =\half\bbra{\rho\uu^{\prime\, 2}}_V, \quad E_{\rm
  kin}^{\rm dif}=\half\bbra{\rho\mean{u_\phi}^2}_V\\ &\text{and}&\ E_{\rm
  kin}^{\rm mer} =
\half\bbra{\rho\left(\mean{u_r}^2+\mean{u_\phi}^2\right)}_V,
\label{eq:ene2}
\end{eqnarray}
where $\bbra{}_V$ indicate a volume average.
In a similar way, the total magnetic energy,
\begin{equation}
E_{\rm mag}^{\rm tot}=\bbra{{\BB^2\over2\mu_0}}_V
\label{eq:ene3}
,\end{equation}
can be decomposed into energies of the fluctuating fields, and the toroidal and poloidal
magnetic fields:
\begin{equation}
E_{\rm mag}^{\rm flu}=\bbra{{\BB^{\prime\,2}\over2\mu_0}}_V, E_{\rm
  mag}^{\rm tor}=\bbra{{\meanB_\phi^2\over2\mu_0}}_V \text{and}\
E_{\rm mag}^{\rm pol}=\bbra{{\meanB_r^2+\meanB_\theta^2\over2\mu_0}}_V.
\label{eq:ene4}
\end{equation}

To determine the turbulent transport coefficients 
from 
our simulations, we use the
test-field method \citep{SRSRC05,SRSRC07,BRRS08,WRTKKB17} with the new convention
introduced in \cite{VKWKR19}. In the test-field method, nine independent
test fields are used to calculate how the flow 
acts on these fields to generate a small-scale magnetic field and therefore an
electromotive force $\EMF$.
It is important to note that these test fields
do not have a feedback effect on the 
simulated hydromagnetic quantities, and are therefore
only diagnostics of the system. 
With the electromotive forces for each test field at hand,
we have a large enough set of equations to solve for
the turbulent transport coefficients from the ansatz
expanding the $\EMF$ in terms of the mean magnetic field
\citep{KR80}:
\begin{equation}
\EMF=\aalpha\cdot\meanBB+\ggamma\times\meanBB 
-\bbeta\cdot(\nab\times\meanBB) 
-\ddelta\times(\nab\times\meanBB) 
-\kkappa \cdot(\nab\meanBB)^{(s)},
\end{equation}
where $(\nab\meanBB)^{(s)}$ is the symmetric part of the diffusion
tensor. Here we have 
neglected contributions of higher-than first-order
derivatives. Also, 
$\aalpha$ and $\bbeta$ are two-rank tensors, $\ggamma$ and
$\ddelta$ are vectors, and $\kkappa$ is a three-rank tensor. These five coefficients
can be associated with 
different turbulent effects important for the magnetic field
evolution: the $\alpha$ effect \citep{SKR66}
can lead to field amplification via helical flows, for example in
convection influenced by rotation; the $\gamma$ effect 
describes turbulent transport of the mean magnetic field in the same way as a mean flow;
 $\beta$ describes turbulent diffusion; and the $\delta$ effect, also
known as the R\"adler 
effect \citep{KHR69}, can lead to dynamo action in the
presence of 
other effects, 
for example the $\alpha$ effect or shear, although it alone cannot lead to the growth
of magnetic energy \citep{BS05}, and the $\kappa$ effect, 
the physical interpretation of which is currently unclear. 
However, the  $\kappa$ effect can contribute, in
theory, to both the amplification and diffusion of magnetic
fields.

In most cases, we express the measured quantities in a nondimensional form by
normalizing them appropriately.
For example, we define
$\alpha_0=\urmsp/3$ and $\etatz=\tau_{\rm tur} u^{\prime\,2}_{\rm rms}/3$
as normalizations for $\aalpha$ tensor, and $\bbeta$ and $\ddelta$ tensors, respectively.
However, sometimes we transfer them into physical units 
by defining the unit system based on
the solar rotation rate $\Omega_\odot=2.7\times10^{-6} \s^{-1}$,  solar
radius $R=7\times10^{8} \m$, density at the bottom
of the convection zone $\rho(0.7R)=200 \kg/\m^3$, and
$\mu_0=4\pi\cdot10^{-7}$~H~m$^{-1}$.
All simulations including the test-field method were performed using the {\sc Pencil
  Code}\footnote{\url{https://github.com/pencil-code/}}.

\section{Results}

All simulations were run to the saturated stage and then
continued with the test-field method switched on for
another 50 to 100 years.
All the computed diagnostic quantities and plots shown
below are obtained from this time interval.
The hydrodynamic (HD) runs are labeled with an `H', and magnetohydrodynamic (MHD) ones with an `M',
while the number in the run label represents these rotation rates normalized
to the solar one, $\Ot$.

We aimed at keeping
all input parameters the same and only changed the rotation
rate as shown in \Tab{runs}, but for some runs this strategy partially failed.
For Runs~H7 and H10, we had to increase the viscosity to stabilize
the simulations against strong shearing motions. 
For Runs~M10 to M30 we
decreased all diffusivities ($\nu,\eta,\chiSm$) 
but kept the
Prandtl numbers ($\PrSGS,\Pm$) 
unchanged in order to have roughly
constant Reynolds numbers ($\Rey$, $\Rm$).

The Rayleigh number 
for Run~M5 is around 100 times
the critical value \citep{WRTKKB17}.
However, the critical Ra is known to increase as a function of
rotation \citep{Ch61}.
Hence, our slower(faster) rotating runs can be expected to exceed the
critical value of the onset of convection by a larger(smaller) margin
than M5.
This might not be the ideal modeling strategy; a better approach would be to fix the
level of supercriticality in each run, but this is currently computationally
too expensive for such a large parameter study.
Furthermore, as verified by
\cite{WRTKKB17} for a run similar to Run~M5, we do not expect a small-scale dynamo to be operating in our simulations.

Finally, referring back to our earlier studies, we note the following.
Runs~M0.5 to M10 have already been discussed in \cite{W18} to
determine the dynamo cycle properties,
but not all the turbulent transport coefficients were presented in that
study.
Run~M5 is similar to Run~I in \cite{WKKB14}, Run~A1 in
\cite{WKKB16}, Run~D3 in \cite{KKOWB17} and \cite{WRTKKB17}, and
Run~G$^{\rm W}$ in \cite{VWKKOCLB17}. Furthermore, Run~M3 is similar to Run~B1 in
\cite{WKKB16} and Runs~M10 and M15 are similar to Runs~I$^{\rm W}$ and J$^{\rm W}$ of
\cite{VWKKOCLB17}.

\begin{figure}[t!]
\begin{center}
\includegraphics[width=0.965\columnwidth]{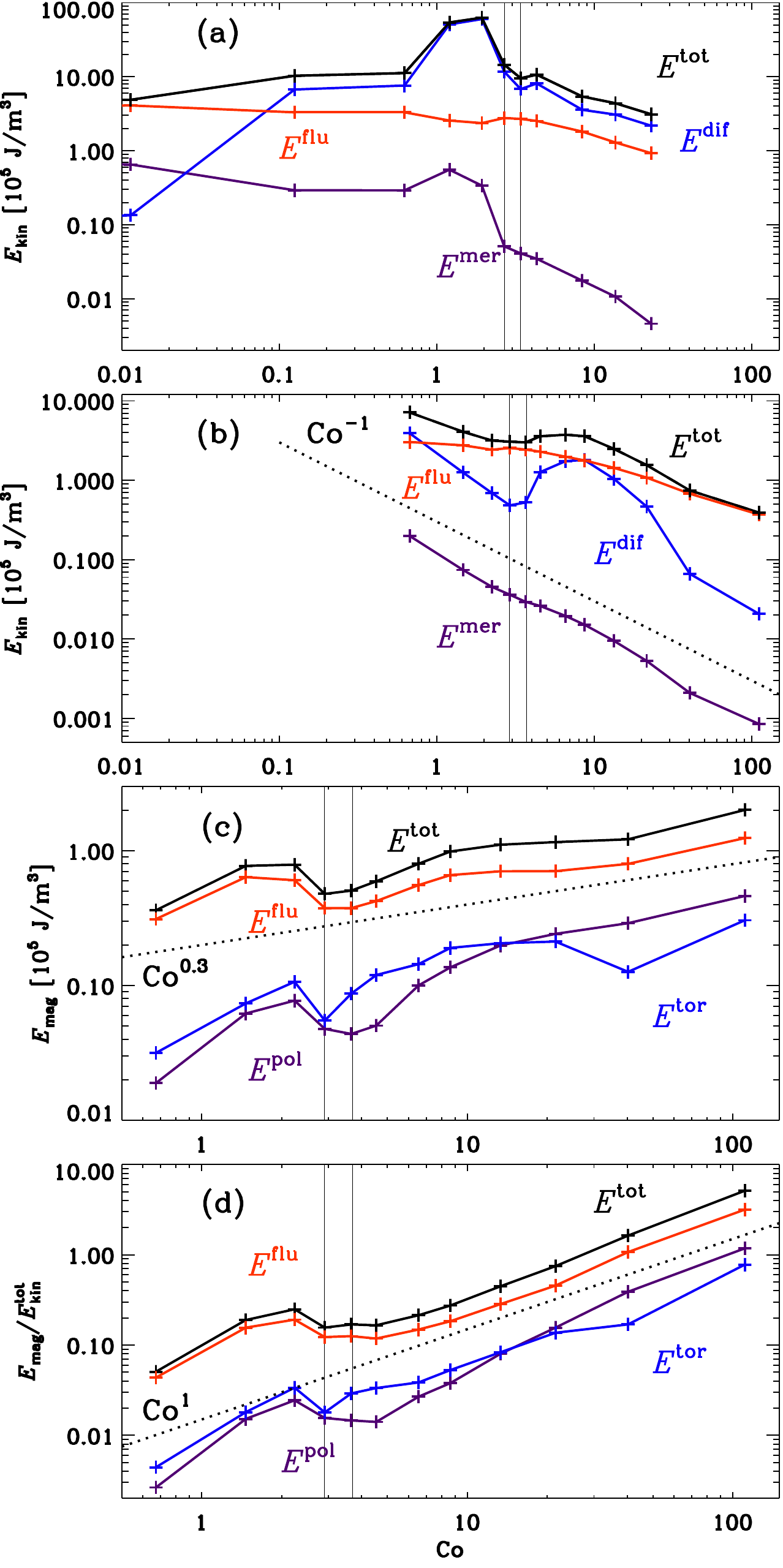}
\end{center}\caption[]{
Dependence of kinetic and magnetic energies on rotation in terms of
Coriolis number $\Co$. We show the total kinetic energy $E_{\rm kin}^{\rm tot}$ (black lines), which is
composed of the energy of the fluctuating flows $E_{\rm kin}^{\rm
  flu}$ (red lines), the differential rotation $E_{\rm kin}^{\rm dif}$ (blue lines), and
the meridional circulation $E_{\rm kin}^{\rm mer}$ (purple lines) for
the HD runs (Set~H) in panel (a) and for MHD runs (Set~M) in panel (b).
Additionally, we show the total magnetic energy $E_{\rm mag}^{\rm
  tot}$ (black line) composed of the energy of the fluctuating
magnetic field $E_{\rm mag}^{\rm flu}$ (red), of the toroidal
$E_{\rm mag}^{\rm tor}$ (blue), and poloidal magnetic field $E_{\rm
  mag}^{\rm pol}$ (purple) in panel (c) and normalized by total kinetic energy $E_{\rm
  mag}^{\rm tot}$ in panel (d). The dotted lines show the following relations
between energies and Coriolis number: $E_{\rm
  kin}\propto\Co^{-1}$ in panel (b), $E_{\rm mag}\propto\Co^{0.3}$ in panel (c), and $E_{\rm mag}/E_{\rm
  kin}\propto\Co^1$ in panel (d).
 The transition from anti-solar to
solar-like differential rotations occurs in between the vertical lines (left line: last anti-solar run,
right line: first solar-like run).
See \Eqss{eq:ene1}{eq:ene4} for the definition of the energies.
}\label{energies}
\end{figure}

\begin{table*}[t!]\caption{
Summary of runs.
}\vspace{12pt}\centerline{\begin{tabular}{l|rrcccr|c|rrrrrrrr}
Run & $\Ot$ &$\Ta$[$10^{6}$]& $\Ra$[$10^{7}$]&$\PrSGS$&$\Rey$&$\Co$&DR&$E^{\rm tot}_{\rm kin}$&$E^{\rm dif}_{\rm kin}$&$E^{\rm mer}_{\rm kin}$&$E^{\rm flu}_{\rm kin}$&$E^{\rm tot}_{\rm mag}$&$E^{\rm tor}_{\rm mag}$&$E^{\rm pol}_{\rm mag}$&$E^{\rm flu}_{\rm mag}$\\[.8mm]   
\hline
\hline\\[-2.5mm]
H0        &  0.0     & 0.0       &  4.0 & 2.0 & 52 & 0.0     &?&  4.863 & 0.127 & 0.706 & 4.031\\
H0.005 &  0.005 & 1.3(-4) &  4.0 & 2.0 & 52 & 0.006  &?&  4.884 & 0.124 & 0.677 & 4.083\\
H0.01   &  0.01   & 5.4(-4) &  4.0 & 2.0 & 52 & 0.011  &AS&  4.873 & 0.134 & 0.652 & 4.087\\
H0.1     &  0.1     & 5.4(-2) &  4.0 & 2.0 & 47 & 0.12    &AS&10.307 & 6.727 & 0.293 & 3.298\\
H0.5     &  0.5     & 1.3       &  4.0 & 2.0 & 47 & 0.62    &AS&11.203 & 7.608 & 0.292 & 3.303\\
H1        &  1.0     & 5.4       &  4.0 & 2.0 & 49 & 1.2      &AS&54.087  &50.976 & 0.554 & 2.557\\
H1.5     &  1.5     & 9.7       &  4.0 & 2.0 & 41 & 1.9      &AS&62.036  &60.103&  0.337 & 2.364\\
H2        &  2.0     &21.6      &  4.0 & 2.0 & 44 & 2.7      &AS&14.515 & 11.699 & 0.051 &2.765\\
H2.5     &  2.5     &33.7      &  4.0 & 2.0 & 43 & 3.4      &S&  9.541 & 6.811 & 0.041 & 2.689\\
H3.0     &  3.0     &48.6      &  4.0 & 2.0 & 41 & 4.3      &S&10.669 & 8.109 & 0.035 & 2.526\\
H5        &  5.0     &125       &  4.0 & 2.0 & 34 & 8.4      &S&  5.402 & 3.569 & 0.018 & 1.815\\
H7        &  7.0     &190       &  3.4 & 2.4 & 26 & 13.6    &S&  4.381 & 3.078 & 0.011 & 1.129\\
H10      & 10.0    &260       &  2.8 & 2.9 & 18 & 22.9    &S&  3.107 & 2.176 & 0.005 & 0.926\\
\hline
M0.5     &  0.5 & 1.3 &  4.0 & 2.0 & 44 & 0.7    &AS& 7.141  & 3.910 & 0.199 &3.032&0.362&0.032&0.019&0.311\\
M1        &  1.0 & 5.4 &  4.0 & 2.0 & 40 & 1.5    &AS& 4.084  & 1.259 & 0.074 &2.751&0.775&0.074&0.063&0.638\\
M1.5     &  1.5 & 12  &  4.0 & 2.0 & 39 & 2.2    &AS& 3.163  & 0.691 & 0.045 &2.427&0.789&0.107&0.077&0.605\\
M2        &  2.0 & 22  &  4.0 & 2.0 & 40 & 2.9    &AS& 3.065  & 0.483 & 0.036 &2.547&0.479&0.055&0.048&0.376\\
M2.5     &  2.5 & 34  &  4.0 & 2.0 & 40 & 3.7    &S& 2.992  & 0.524 & 0.029 &2.438&0.506&0.087&0.044&0.375\\
M3        &  3.0 & 49  &  4.0 & 2.0 & 39 & 4.5    &S& 3.584  & 1.268 & 0.026 &2.290&0.593&0.120&0.050&0.423\\
M4        &  4.0 & 86  &  4.0 & 2.0 & 36 & 6.6    &S& 3.741  & 1.741 & 0.019 &1.981&0.801&0.144&0.100&0.557\\
M5        &  5.0 & 35  &  4.0 & 2.0 & 34 & 8.6    &S& 3.600  & 1.804 & 0.015 &1.780&0.987&0.190&0.136&0.660\\
M7        &  7.0 & 264 &  4.0 & 2.0 & 31 & 13.4  &S& 2.481  & 1.040 & 0.009 &1.432&1.109&0.206&0.198&0.704\\
M10      & 10.0& 540 &  4.0 & 2.0 & 27 & 21.5  &S& 1.550  & 0.465 & 0.005 &1.079&1.159&0.212&0.242&0.705\\
M15      & 15.0& 1897 &  7.4 & 2.0 & 27 &40.3  &S& 0.746  & 0.066 & 0.002 &0.677&1.216&0.126&0.290&0.799\\
M30      & 30.0& 13488 & 16.1 & 2.0 & 26 &110.9  &S& 0.392  & 0.021 & 0.001 &0.370&2.007&0.305&0.462&1.241\\
\hline
\hline
\label{runs}\end{tabular}}\tablefoot{
Second to fourth columns: input parameters. Columns 6 to 15
show the output parameters, which are calculated from the saturated
stage of the simulations. DR indicates the type of differential
rotation, either it is anti-solar (AR), or solar (S)-like or
inconclusive (?) differential
rotation. The energies $E$ are given in $10^5$ J/m$^2$
and their definitions are given in \Eqss{eq:ene1}{eq:ene4}. All runs
have a density contrast of
$\Gamma_\rho\equiv\rho(r=0.7R)/\rho(R)=31$ and the MHD runs (Set~M) $\Pm=1$. The resolution is $180\times256\times128$ grid points  for all
runs, except it is $360\times512\times256$
for Run~H10.
}
\end{table*}

\subsection{Rotational influence on kinetic and magnetic energies}

First, we discuss how the different 
energies are
influenced by rotation.
As shown in \Fig{energies}(a) and (b), the total kinetic energy increases for
slow rotation, with a maximum for the runs with strong anti-solar
differential rotation. 
For rapidly rotating cases, the kinetic energy drops strongly because of the
rotational quenching of convection.
For the MHD runs, we do not find a maximum during the anti-solar differential
rotation phase, but rather exhibit a dip during the 
transition; otherwise they also fall off, as in the HD runs.
For the HD runs, the kinetic energy is dominated by the
differential rotation, the fluctuating fields
contribute around 10-30\% for most of the runs, and the contribution
of the meridional circulation is weak for all runs, in particular for
high rotation. For MHD runs, the contribution of the fluctuating
field is dominating, and energy related to differential rotation and
meridional circulation becomes even weaker with higher rotation.
This is consistent with the findings of \cite{VWKKOCLB17}, where
the energy of the fluctuating field is  also dominating the kinetic energy
for most of the runs.
This is why we see that relaxing the wedge assumption does not strongly
influence the energy balance in the flow field itself; the most dominant
factor is the inclusion or exclusion of the magnetic fields.
In the MHD cases, the energy of the meridional and differential
rotation decreases roughly linearly overall, whereas the energy of the
fluctuating and total fields decreases with a less steep slope.

All magnetic energy contributions show a weak increase with
rotation, as shown in \Fig{energies}c.
The fluctuating field is also dominating the total energy here, whereas the 
contribution from mean fields ($E_{\rm mag}^{\rm tor}+E_{\rm mag}^{\rm
  pol}$) increases from 10\% for slow rotation
to around 40\%  of the total magnetic energy for the highest rotation. 
This is mostly because the poloidal contribution becomes stronger for
larger rotation, whereas the toroidal 
contribution remains roughly constant.
Also the magnetic energy shows a small enhancement for the runs with
anti-solar differential rotation, however in this case it is only one
run.
This seems to be in contradiction to the work by \cite{BG18}, where
they re-plotted the data of \cite{KKKBOP2015}, finding a stronger
increase with decreasing Co than in the present study. However, in the
original work of \cite{KKKBOP2015}, the change in the magnetic field
as a function of rotation appears insignificant with respect to the large error bars.
However, we note that there is a difference between how the Co is
changed in \cite{KKKBOP2015} and how it changes in our study: these
latter authors change the thermal Prandtl number and therefore the strength of
convection, in which case their Taylor number did not change, in
contrast to our strategy.
Furthermore, the setup of these latter authors is different from ours in terms of 
the radial extent. 
These differences also lead to a transition from anti-solar to
solar-like differential rotation at a quite different Co.

The ratio of magnetic energy to kinetic energy
gives an indication for the dynamo efficiency for each run; see \Fig{energies}d.
We find that this ratio increases roughly linearly with rotational influence on covection, measured by $\Co$.
However, this is not only due to the fact that the kinetic
energy decreases; the increase in magnetic energy  as seen
in \Fig{energies}c also plays a role.
However, the decrease in kinetic energy seems to be the dominant
behavior here. 
This is consistent with the findings of \cite{VWKKOCLB17}, where their Figure
8 shows a linear tendency, but with a larger spread, and the work of
\cite{ABT19}, in which the authors collect data from a variety of
simulations and  also find a linear dependency with $\Co$.
We note here that the MHD runs are probably a more realistic
representation of stars, as real stars have magnetic fields with
similar strengths to those in our models. Hence, the dominance of the energy
contribution of the differential rotation in HD, in particular in the
anti-solar differential rotation cases, might be an artifact.

Comparing these modeling results with observations of stellar magnetic
activity reveals fundamental differences.
X-ray luminosity \citep[e.g.,][]{PMMSBV03,WD16}, Ca II
H\&K emission \citep[e.g.,][]{BST1998}, and surface magnetic field measurements
using either Zeeman-Doppler imaging (ZDI)
\citep[e.g.,][]{VGJDPMFBC14} or Doppler imaging \citep[e.g.,][]{Sa01}
show an increase with Co with a power of around one or two for
$\Co\lesssim 10$.
Our models show a weak increase of the magnetic energy with a power of
around 0.3 over all runs; see \Fig{energies}c. This corresponds to an
increase of around 0.15 in terms of the magnetic field strengths.
This is in conflict with the observational results in two
respects. First of all, the increase of magnetic field strength with Co is  too low to
be comparable with observations by a factor of between four and ten.
Second, our models do not reproduce any two-dependency behavior, where
for slow and moderate rotation ($\Co\lesssim 10$) the magnetic field
increases with rotation and for rapid rotation ($\Co\gtrsim 10$) the
energy is independent of rotation. Our results are only consistent
with the second behavior as our magnetic energy is only weakly
increasing with Co. However, our models show this behavior already for
slow and moderate rotation.
One of the reasons for this discrepancy is the rotational dependence of
supercriticality of convection \citep{Ch61}. Because our Rayleigh number remains mostly
constant, the convection is highly quenched by rotation as seen in
the rotational dependence of the kinetic energy; see \Fig{energies}b.
In real stars, the supercriticality is so high that the rotational
quenching will not be important. Hence, the kinetic energy will be
most likely independent of rotation. However, even if we take this
into account and use the ratio of kinetic to magnetic energy as
a function of Co to calculate the increase in magnetic field strength
with Co, our models are still inaccurate by a factor of between two and four.

Furthermore, we want to note that even if the magnetic field strength 
increases only mildly as a function of rotation,
its surface topology can influence 
coronal X-ray emission. Higher rotation leads to more helical fields,
as also shown in \Sec{sec:alpha_rot}, and this 
in turn can lead to
higher X-ray luminosity as shown by \cite{WP20}.
Therefore, a weak increase in magnetic field strength might 
not be in contradiction with
the strong increase in X-ray luminosity as a function of
rotation for small and moderate rotation.

\begin{figure*}[t!]
\begin{center}
\includegraphics[width=\textwidth]{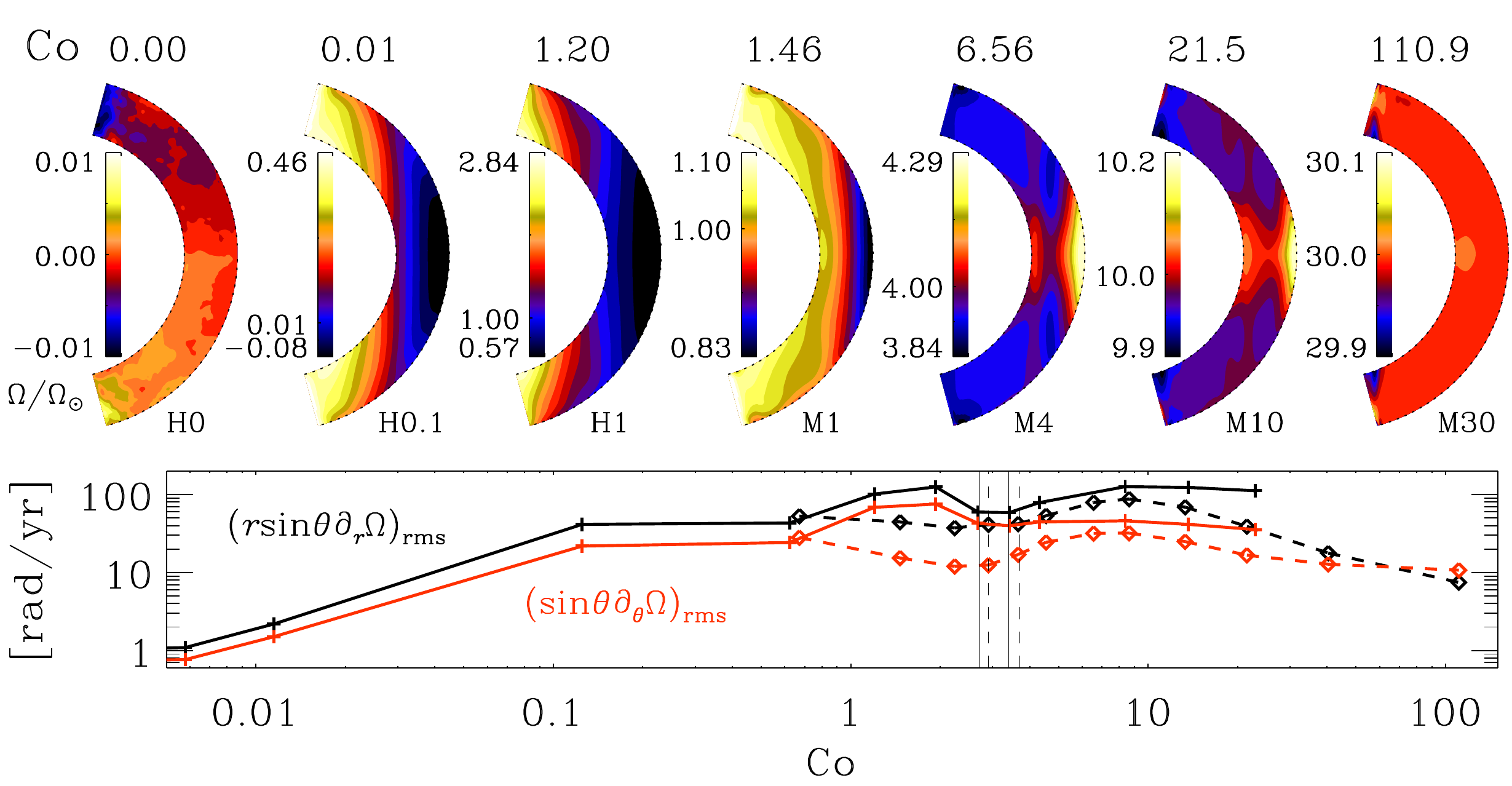}
\end{center}\caption[]{
Normalized local rotation profiles $\Omega/\Omega_\odot$ with
$\Omega=\Omega_0+\mean{u_\phi}/r\sin\theta$ for Runs~H0, H0.1,
H1, M1, M4, M10, M30 and the rms values of the radial
$r\sin\theta\partial\Omega/\partial r$ (black line) and latitudinal
shear $\sin\partial\Omega/\partial \theta$ (red) versus Coriolis number
$\Co$. The values have been calculated as a time average over the
saturated state, and we have omitted the 5 closest grid points to the
latitudinal boundary to remove 
boundary effects from the rms.
The values of the MHD runs (Set~M) are shown with a
dashed line with diamonds.
The zero rotation run has been moved to $\Co=10^{-4}$ to be visible in the lower panel.
The transition from anti-solar to
solar-like differential rotations occurs in between the vertical lines  (left line: last anti-solar run,
right line: first solar-like run) and is shown as solid (Set~H) and dashed (Set~M) lines.
}\label{diffrot}
\end{figure*}

\subsection{Differential rotation and shear}
\label{sec:diff}

As already mentioned in the previous section, the differential
rotation is strongly affected by rotation.
As shown in \Fig{diffrot}, for no or very slow rotation (Runs~H0 and
H0.005) the
differential rotation is very weak showing an inclusive pattern; see
also \Tab{runs} and an overview of all differential rotation profiles
in \Figs{diffrot_HD}{diffrot_MHD}.
For slow to moderate rotation ($\Ot=0.01$ to $2$), we find anti-solar
differential rotation.
For runs with higher rotation, this switches to solar-like differential
rotation, where the equator rotates faster than the poles.
This transition   has already  been found in previous studies of stellar
and planetary dynamos
\citep{GYMRW14,KMB14,FF14,FM15,KKKBOP2015,VWKKOCLB17}.
For rapidly rotating runs, the differential rotation becomes very weak
and mostly pronounced at the equator and at the latitudinal
boundaries.

As we are mostly interested in the analysis of the dynamo drivers in
these runs, we focus next on how the shear acting of the magnetic field
changes with rotation rate.
As shown in the lower panel of \Fig{diffrot}, the rms value of the latitudinal and
radial shear averaged over the whole convection zone increases for
slow rotation when the differential rotation builds up.
For the HD runs (Set~H), this value has a maximum for the anti-solar differential
rotation runs, and decreases during the differential rotation
transition. The radial differential rotation also has a maximum in the
solar-like differential rotation regime, while the latitudinal one
remain roughly constant.
For all HD runs, the radial differential rotation 
is 1.5 to 3 times larger than the latitudinal one, where the largest
differences are seen for the rapidly rotating runs.
For the MHD runs (Set~M), the latitudinal shear is roughly independent
from rotation rate, while the radial shear increase for the solar-like
differential rotation cases and decreases for rapidly rotating runs.
Also, for the MHD runs the radial shear is stronger than the
latitudinal one, except for the runs with $\Co=20$ and larger.
Hence, the weak increase of magnetic energy with rotation cannot be
explained by an increase in shear, as the shear either remains constant
or declines for large rotation rates.

\begin{figure*}[t!]
\begin{center}
\includegraphics[width=\textwidth]{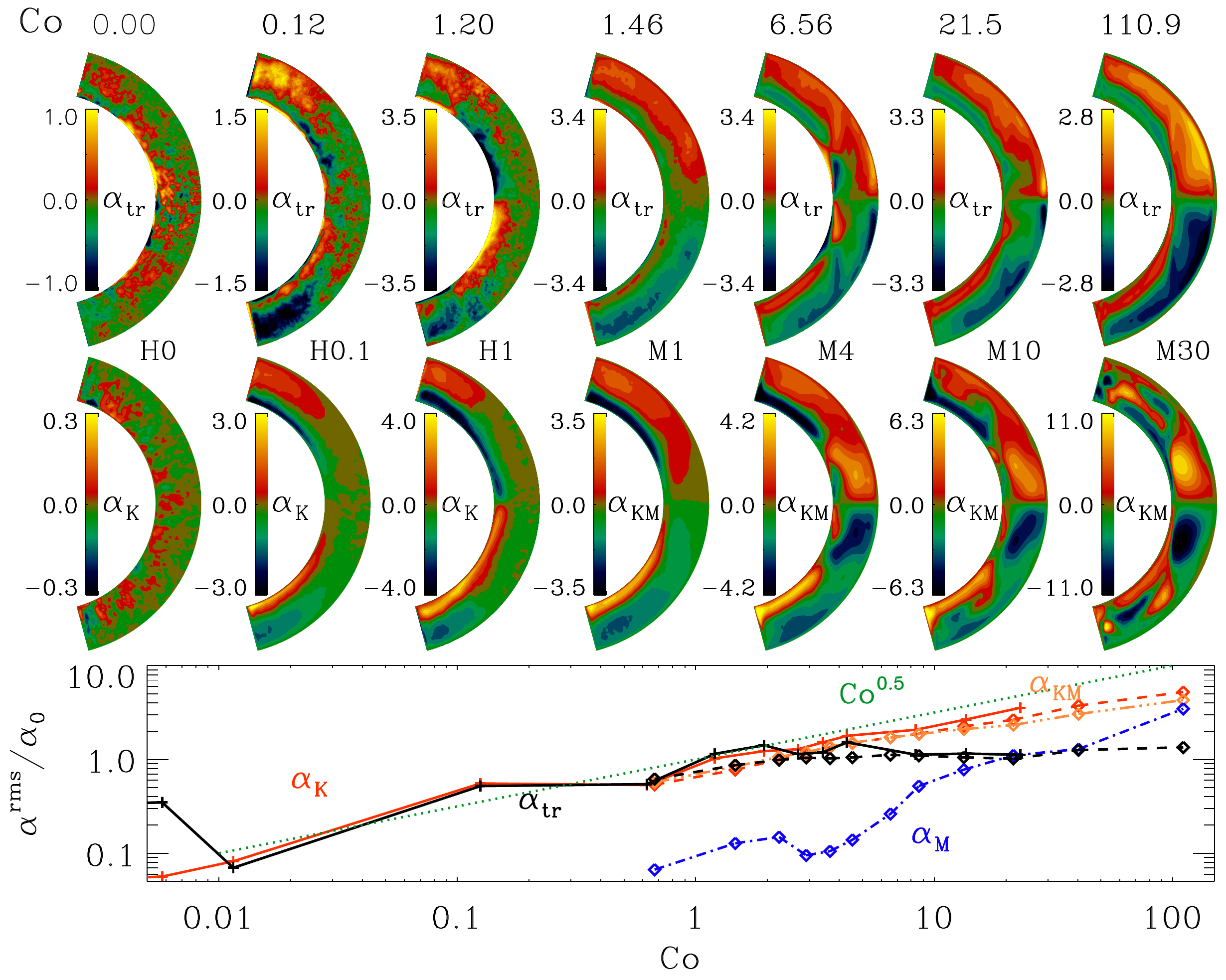}
\end{center}\caption[]{
Rotational dependence of the $\aalpha$ profiles calculated with test-field
method and calculated from helicities.
In the first row, we plot the traces of the $\aalpha$ tensors,
$\alpha_{\rm tr}$, which were determined using the test-field method for a selection of HD (Set~H) and
MHD runs (Set~M). In the second row, we plot the kinetic $\alpha_{\rm K}$ and the sum
of kinetic and magnetic alpha $\alpha_{\rm KM}$ for the corresponding
runs of Set~H and Set~M, respectively.
In the last row, we show the root-mean-square values of the trace of
the tensor $\alpha_{\rm tra}$ (black lines), the kinetic $\alphaK$ (red),
the magnetic $\alphaM$ (blue), and their sum $\alpha_{\rm KM}$
(orange) for all runs. The values of the HD runs are shown with a solid line and crosses, whereas the MHD runs are shown with a
dashed line with diamonds. The greed dotted line indicates a power law
with an exponent of $0.5$. All values are normalized by
$\alpha_0=\urmsp/3$. The zero rotation run has been moved to
$\Co=10^{-4}$ to be visible in the lower panel.
}\label{alpha_trace}
\end{figure*}

\begin{figure}[t!]
\begin{center}
\includegraphics[width=0.9\columnwidth]{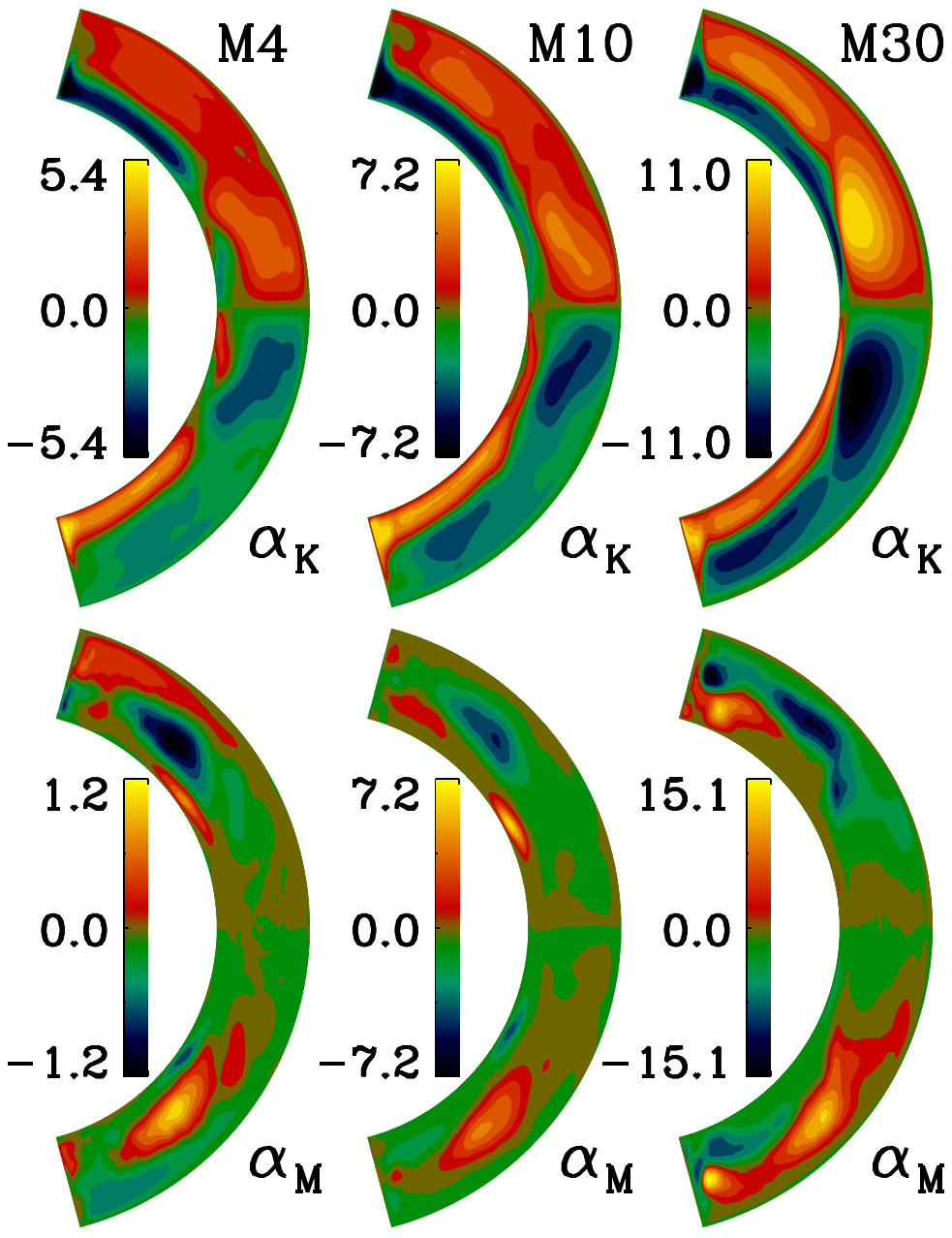}
\end{center}\caption[]{
Kinetic $\alphaK$ and magnetic $\alphaM$ for three of the runs
shown in \Fig{alpha_trace}, which have a significant
contribution from $\alphaM$. All values are normalized by $\alpha_0=\urmsp/3$.
}\label{alpKM}
\end{figure}

\begin{figure}[t!]
\begin{center}
\includegraphics[width=0.9\columnwidth]{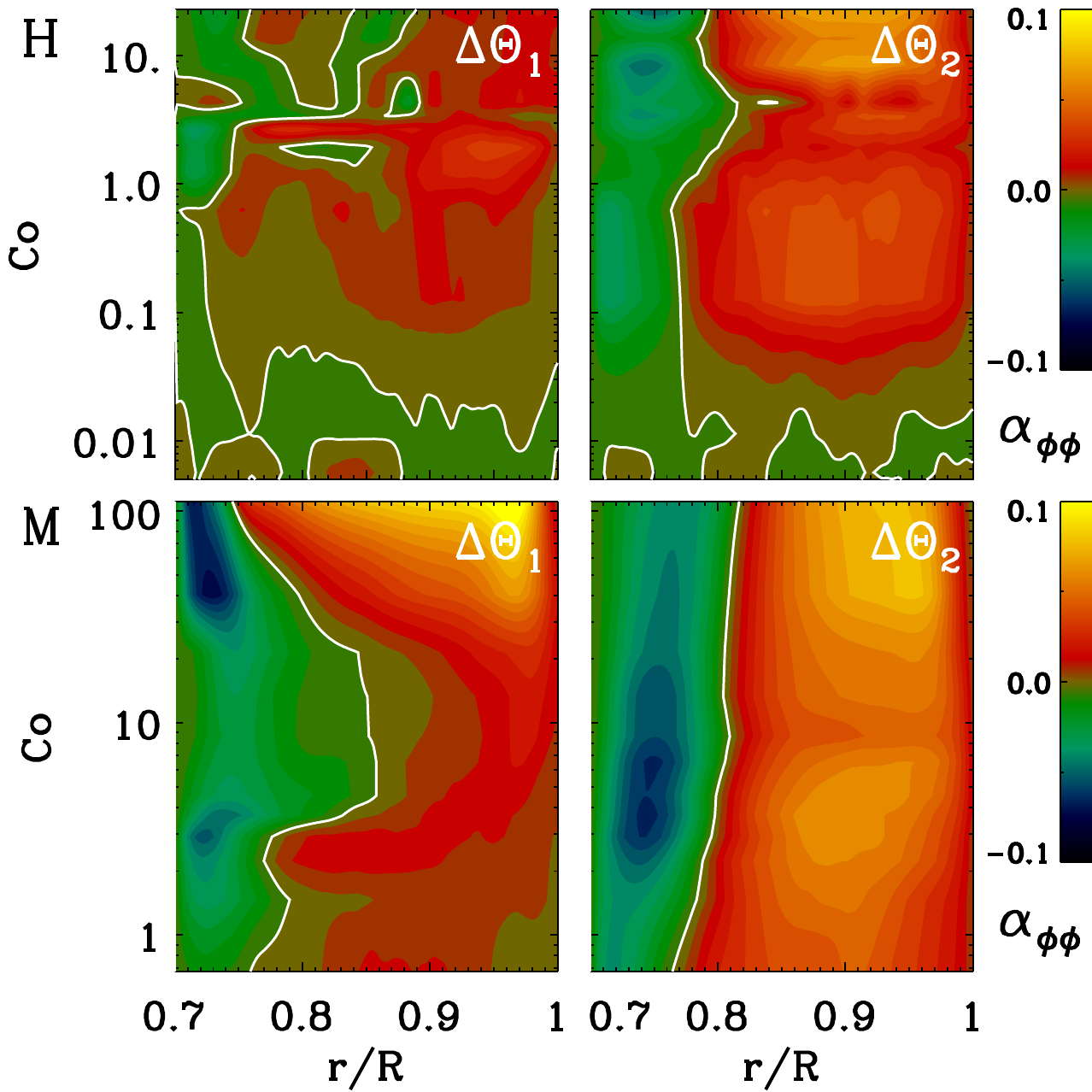}
\end{center}\caption[]{
Radial inversion of $\alphapp$ as a function of $\Co$
for two latitudinal strips. In the first column we average $\alphapp$
 at low latitudes, i.e., $\Delta\Theta_1=10^\circ-20^\circ$  , and
in the second column we average over mid latitudes $\Delta\Theta_2=50^\circ-60^\circ$.
The zero values are indicated with a white line.
All values are normalized by $\alpha_0=\urmsp/3$.
}\label{alphpp_rot}
\end{figure}

\subsection{Rotational dependencies of  $\aalpha$}
\label{sec:alpha_rot}

Now, we investigate the rotational dependence of $\aalpha$ and 
first focus on its general properties. 
We compute the trace
of the $\aalpha$ tensor, $\alpha_{\rm tr}=\alpharr+\alphatt+\alphapp$
using the test-field method.
For comparison, we also calculate $\alpha$ based on the kinetic and
current helicity, following \cite{SKR66} and \cite{PFL76}:
\begin{equation}
\alphaK=-{\tau_{\rm cor}\over
  3}\overline{\fluc{\oo}\cdot\flucuu},\quad \alphaM=  {\tau_{\rm cor}\over
  3}\overline{\fluc{\jj}\cdot\flucbb}/\meanrho, \quad
\alphaKM\equiv\alphaK+\alphaM
\label{eq:aKaM}
,\end{equation}
where $\alphaK$ and $\alphaM$ are the kinetic and magnetic $\alpha$
coefficients, respectively, $\fluc{\oo}=\nab\times\flucuu$ is the
fluctuating vorticity, $\overline{\fluc{\oo}\cdot\flucuu}$ is the small-scale
kinetic helicity, $\fluc{\jj}=\nab\times\flucbb/\mu_0$ is the
fluctuating current density,
$\overline{\fluc{\jj}\cdot\flucbb}$ is the small-scale current
helicity, $\meanrho$ is the mean density, and $\tau_{\rm cor}$ is the
turbulent correlation time, which
we now set equal to the convective turn-over time,
$\tau_{\rm cor}=\tau_{\rm tur}$.
In \Fig{alpha_trace}, we show the meridional profiles and the rms
values of the $\alpha_{\rm tr}$ 
computed both from the HD (Set~H) and MHD (Set~M) runs,
together with $\alphaK$ or $\alphaKM$,
respectively, as a function of $\Co$.
We find that $\alpha_{\rm tr}$ from HD runs closely
follows $\alphaK$ for slow
and moderate rotation ($\Co=0$ to 4) in distribution and amplitude.
$\alpha_{\rm tr}$ from MHD runs is somewhat weaker than the
other quantities.
However, all the different measurements show very similar spatial distributions,
and all show growth consistent with $\Co^{0.5}$.
The magnetic part, $\alphaM$, is an order of magnitude weaker than $\alphaK$,
but is growing with a similar power law as a function of rotation.
$\alpha_{\rm tr}$, $\alphaK$ and $\alphaKM$ show profiles that are positive(negative) in the northern
hemisphere in the upper(lower) part of the convection zone; the signs
are opposite in the southern hemisphere.
This spatial pattern seems to be roughly independent of rotation for
$\alpha_{\rm tr}$s in this moderate rotation regime.

For higher rotation, $\alpha_{\rm tr}$ and $\alphaKM$ significantly decouple, and
the HD and MHD test-field results start following each other tightly.
While $\alphaKM$ and $\alphaK$ still continue growing with the same
power law as in the moderate rotation regime, both the test-field measured
quantities no longer depend on rotation. Here, $\alphaM$ shows a much
stronger dependence on rotation ($\propto \Co$) than any of the other quantities, and becomes
comparable to $\alphaK$ for the most rapidly rotating case.
This difference in between the theoretical prediction and test-field measurements
could be explained by us not modeling the rotational 
dependence of $\tau_{\rm cor}=\tau_{\rm tur}$
correctly. 
The convective scale entering the calculation of $\tau_{\rm tur}$ is
known to be dependent on rotation.
The theoretical calculation of \cite{Ch61} predicts a dependence of $\Co^{-0.3}$, 
while the models of \cite{FH16} and \cite{VWKKOCLB17} show a dependence of $\Co^{-0.5}$ .
If we take this into account, the increase of $\alphaKM$ mostly vanishes.

Also the strongly growing $\alphaM$ contributes to the increase of $\alphaKM$.
In our most rapidly rotating runs $\alphaM$ can, locally, even exceed the value of 
$\alphaK$, as is evident from \Fig{alpKM}.
We see that $\alphaM$ is mostly negative (positive) in
the northern (southern) hemisphere in the upper part of the convection
zone, and positive (negative) below, and therefore it has the
  opposite sign compared to $\alphaK$. The peak values of $\alphaM$ are larger than
$\alphaK$ for rapidly rotating runs, but these 
locations are not those where $\alphaK$ is the strongest.
This leads to a more
complicated distribution of $\alphaKM$, where at high latitudes in the
middle of the convection zone, the sign of $\alphaKM$ changes due to
$\alphaM$, but at low latitudes $\alphaKM$ is still dominated by
$\alphaK$.
Hence, in the most rapidly rotating cases, the $\alphaKM$ profiles no longer closely match with the test-field measured profiles.
The formula of \Eq{eq:aKaM} has been introduced by \cite{PFL76} for
cases where $\alphaM$ is small and acts as a perturbation to
$\alphaK$. In our case, in contrast, $\alphaM$ is even stronger than
$\alphaK$ at some locations, so 
we cannot expect this expression to be valid in this regime.
One possible inconsistency in our approach is to regard
the correlation times, $\tau_{\rm cor}$, of the kinetic and magnetic parts
of the $\alpha$ effect as equal. In reality, this might not
be the case, and our analysis should be refined. In any
case, our current results show that $\alphaKM$, using the procedure
adopted here and very commonly by other authors analyzing their
MHD simulations, 
should be only used as a proxy of the $\alpha$ effect with some caution.

In summary, we find from our simulations that
quenching of the $\alpha$
effect in terms of rotation can be mostly explained by the changes in
the turbulent correlation length.
In addition, we note that,
due to the increasing ratio of magnetic to kinetic energy as a function of rotation, 
we may also be seeing  magnetic quenching reducing the $\alpha$ effect.
By comparing the HD and MHD test-field measurements, however,
we obtain very similar rms values in the regime where the magnetic
field should
be dominant. 
Hence, the magnetic quenching seems to be weak in these
particular runs. 
However, even if the differences in the rms values can be small,
locally there can be strong differences between the HD and the MHD
runs, as found by \cite{WRTKKB17}.

According to the Parker-Yoshimura rule \citep{P55,Yos75}, an 
$\alpha \Omega$ dynamo 
will produce an equatorward migrating dynamo wave if $\alphapp$
and radial shear have different signs on opposite hemispheres. 
Typically, convection simulations produce positive $\alphapp$
in the north, while the radial gradient is weak and positive in the bulk
of the convection zone. There is often a narrow layer of reversed
sign of $\alphapp$ in the bottom of the convection zone, as is also the case in the simulations
presented here, but this is not large enough to contribute to the correct
migratory properties of the wave. Instead, equatorward migration is driven
by an additional local region of negative radial shear together with the 
positive $\alphapp$ in the bulk. Only in the 
thicker shell simulations in the planetary context has there been success
in producing a thick-enough layer of reversed sign of helicity
to drive the equatorward dynamo wave with a positive radial gradient of shear \citep{DWBF15}.
To investigate how the thickness of the inversion layer changes as a function
of rotation, 
we plot in \Fig{alphpp_rot} the radial distribution of
$\alphapp$ for two latitudinal bands at low
($\Delta\Theta_1=10^\circ-20^\circ$)  and
mid latitudes ($\Delta\Theta_2=50^\circ-60^\circ$).
At low latitudes, the 
HD runs do not show an inversion layer at all,
but at the higher latitude band, an inversion layer extending roughly
one fourth of the convection zone is visible, becoming somewhat wider
as a function of rotation. However, the magnitude of $\alphapp$ in this layer 
is very weak.
In the MHD runs, $\alphapp$ is always negative in the lower fourth of the
convection zone, but
for values of $\Co$ of between 4 and 30 this region reaches up to half of
the convection zone at low latitudes.
Increasing rotation even more, the inversion layer again becomes narrower.
For the mid latitudes, the region of negative $\alphapp$ is also
located in the lower third of the convection zone. We find a tendency
for this region to increase for larger rotation.
Here, $\alphapp$ is stronger in the inversion layer in the MHD cases than in the HD ones. 
Hence, we do not find inversion layers extending close to the surface
from our simulations, as is found by \cite{DWBF15}.

\begin{figure*}[t!]
\begin{center}
\includegraphics[width=\textwidth]{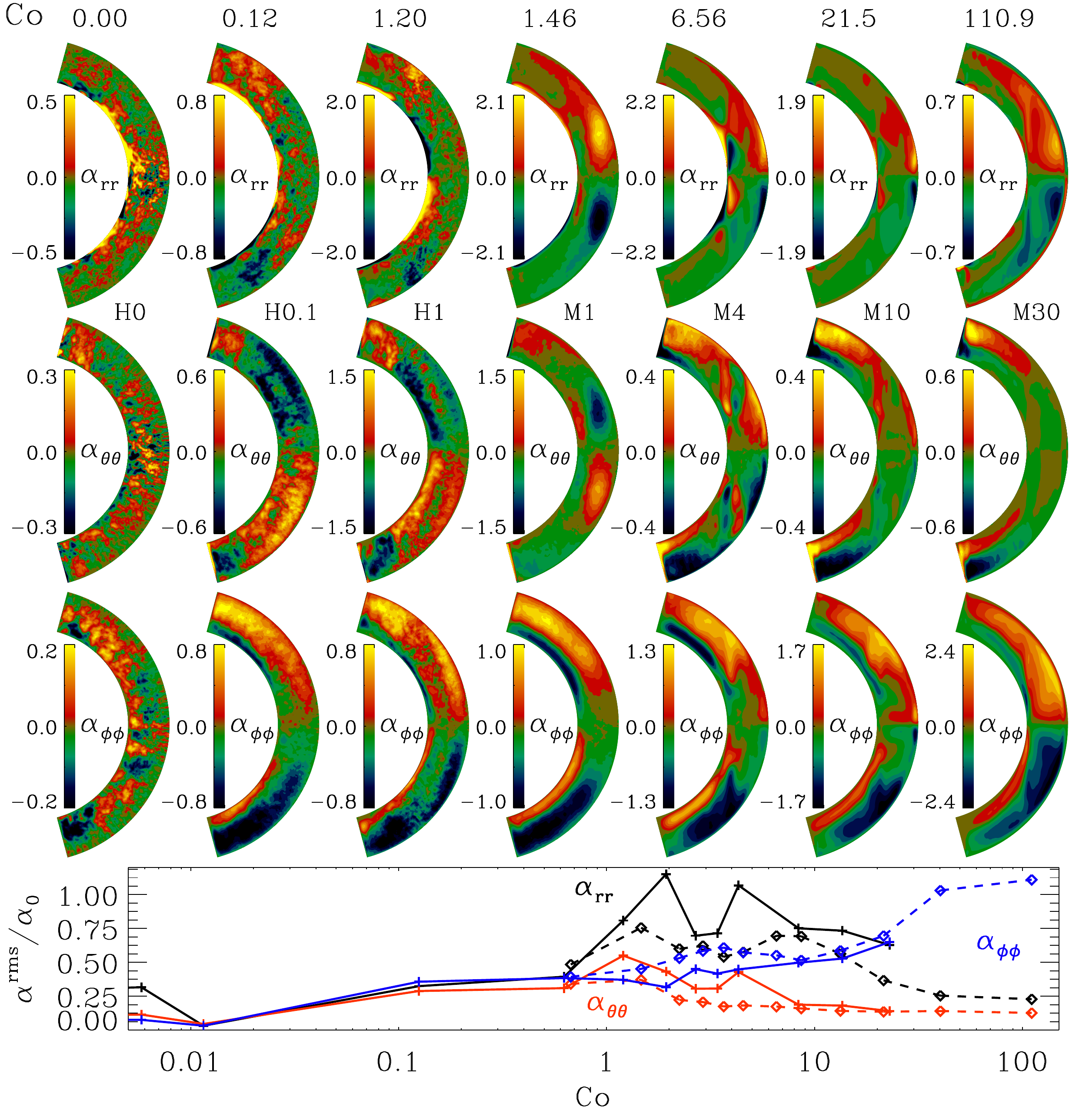}
\end{center}\caption[]{
Rotational dependency of the diagonal $\aalpha$ components. We show
meridional profiles of $\alpharr$ (top row), $\alphatt$ (second row), and
$\alphapp$ (third row), and their rms values (bottom row), with
$\alpharr$ (black lines), $\alphatt$ (red), and $\alphapp$ (blue). 
As in \Fig{alpha_trace} the values of the HD runs (Set~H) are shown
with a solid line and crosses, whereas the MHD runs (Set~M) are shown
with a dashed line with diamonds.
All values are normalized by $\alpha_0=\urmsp/3$.
The zero rotation run has been moved to $\Co=10^{-4}$ to be visible in the lower panel.
}\label{alpha_ani}
\end{figure*}

\subsection{Anisotropy of the $\aalpha$ tensor}

\begin{figure*}[t!]
\begin{center}
\includegraphics[width=\textwidth]{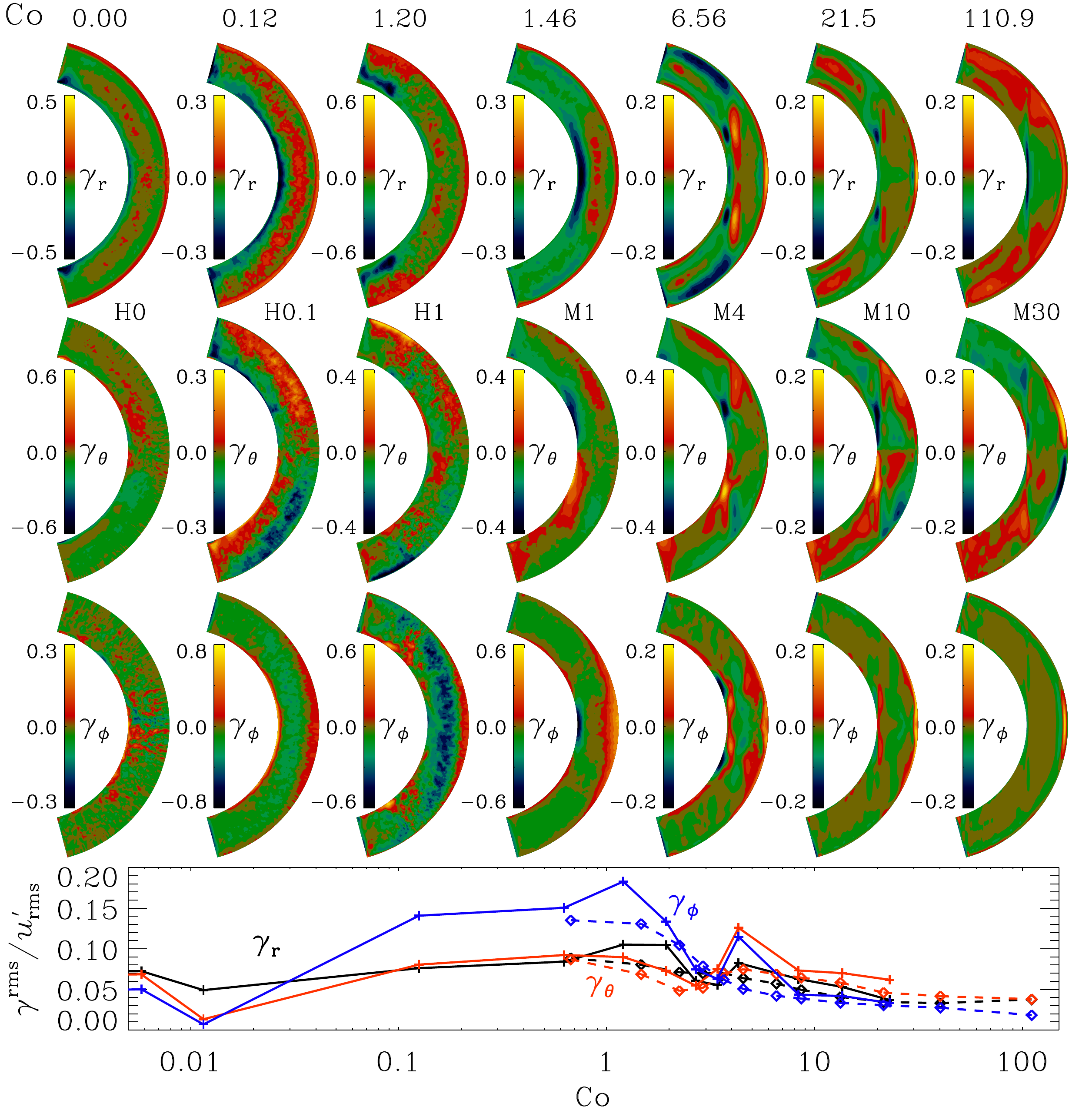}
\end{center}\caption[]{
Rotational dependency of the $\ggamma$ components. We show
meridional profiles of $\gamma_r$ (top row), $\gamma_{\theta}$ (second row), and
$\gamma_{\phi}$ (third row), and their rms values (bottom row), with
$\gamma_r$ (black lines), $\gamma_{\theta}$ (red), and $\gamma_{\phi}$
(blue). 
As in \Fig{alpha_trace}, the values of the HD runs (Set~H) are shown
with a solid line and crosses, whereas the MHD runs (Set~M) are shown
with a dashed line with diamonds.
All values are normalized by $\urmsp$. The
zero rotation run has been moved to $\Co=10^{-4}$ to be visible in
the lower panel.
}\label{gamma}
\end{figure*}

\begin{figure*}[t!]
\begin{center}
\includegraphics[width=\columnwidth]{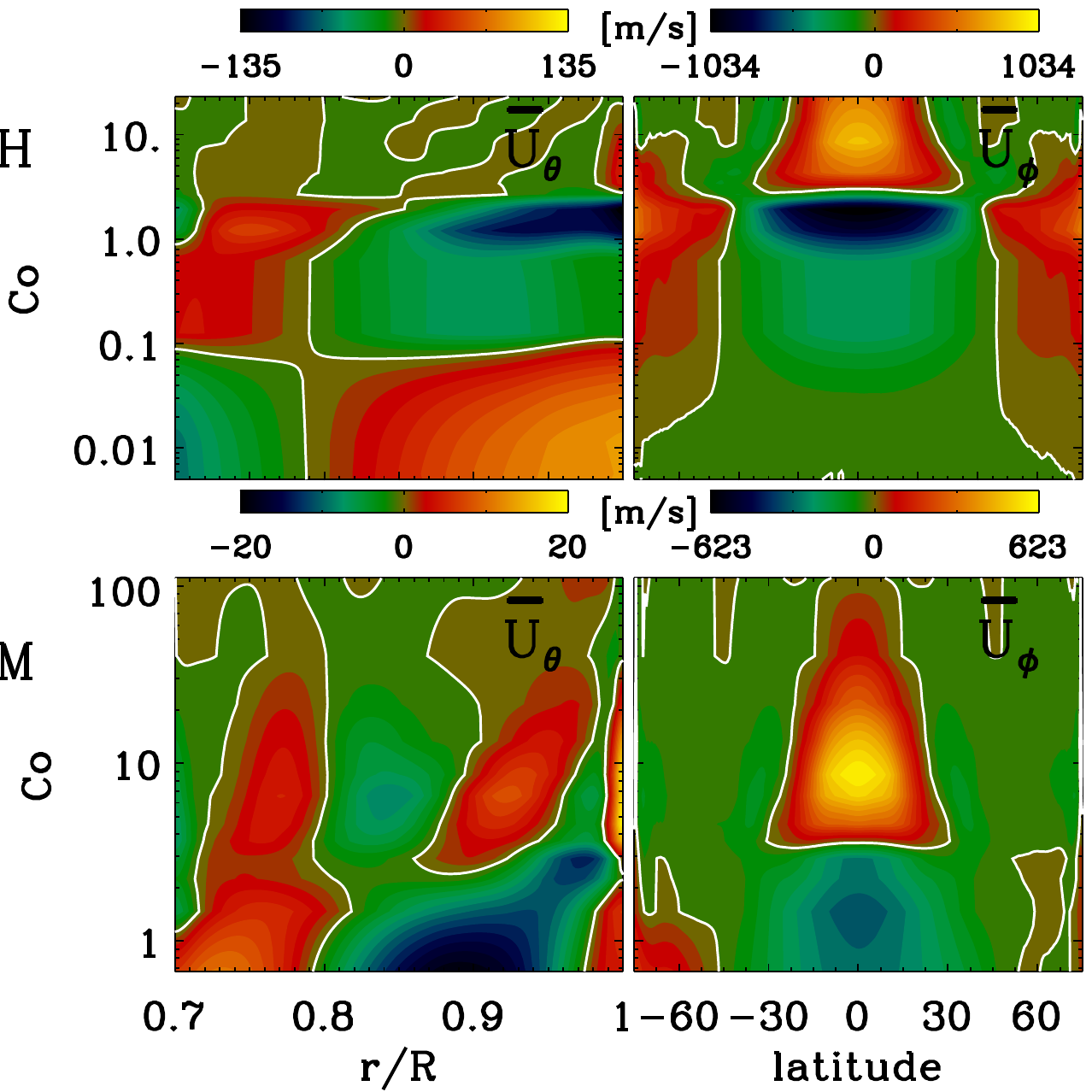}
\includegraphics[width=\columnwidth]{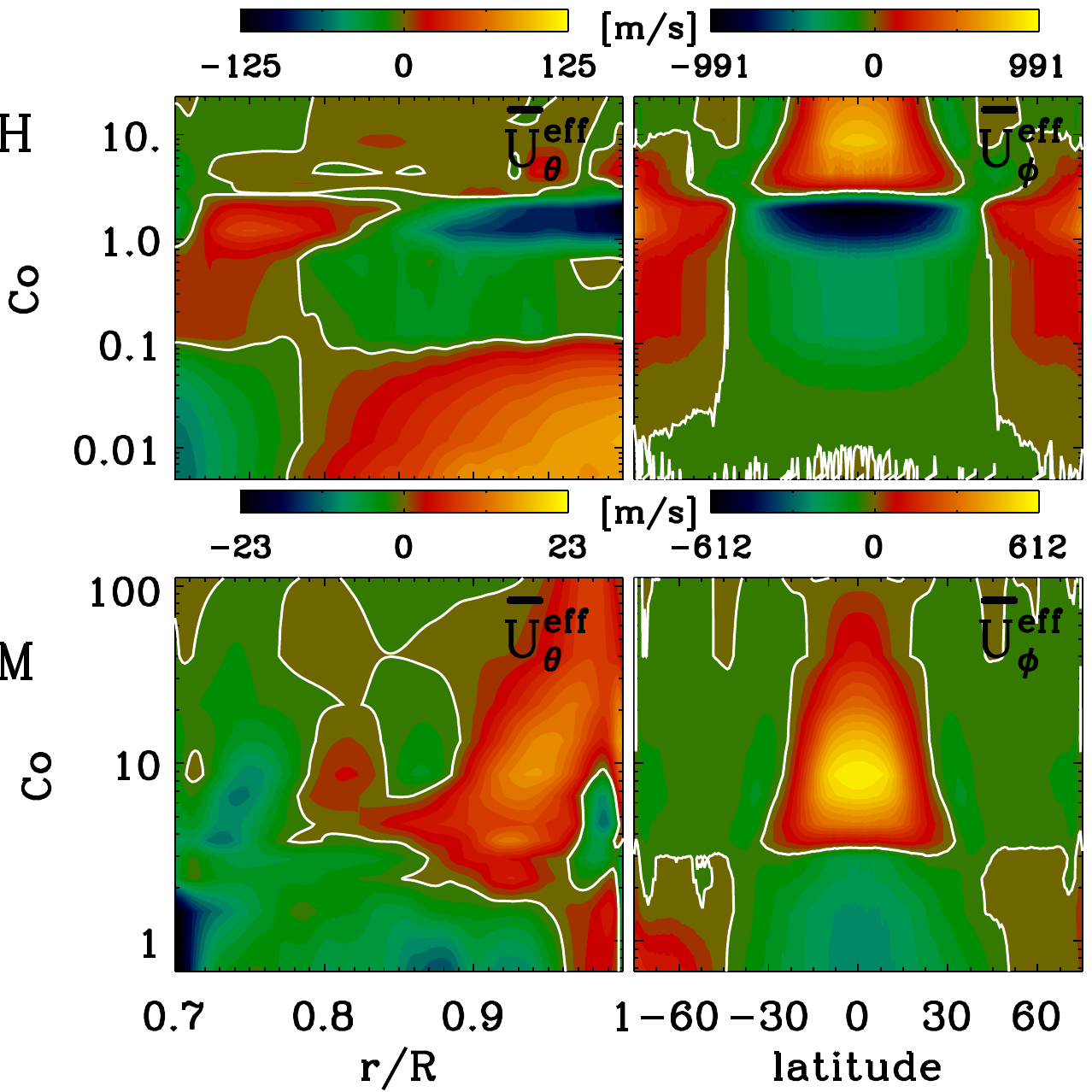}
\end{center}\caption[]{
Rotational dependency of meridional circulation and differential
rotation and their effective counterparts. In the first column we plot the meridional circulation
$\meanU_\theta$ and $\meanU^{\rm eff}_\theta$ at 25$^\circ$ latitude
together with the azimuthal mean velocity $\meanU_\phi$ and
$\meanU^{\rm eff}_\phi$ close to the surface $r=0.99\,R$ as a function
$\Co$ at for the HD runs (Set~H) in the top row and
for the MHD runs (Set~M) in the bottom row. 
Positive (negative) values of $\meanU_\theta$ are equatorward
(poleward) and positive  (negative) $\meanU_\phi$ are prograde
(retrograde).
The zero
values are indicated with a white line.
}\label{meri+diff_rot_eff}
\end{figure*}

\begin{figure}[t!]
\begin{center}
\includegraphics[width=\columnwidth]{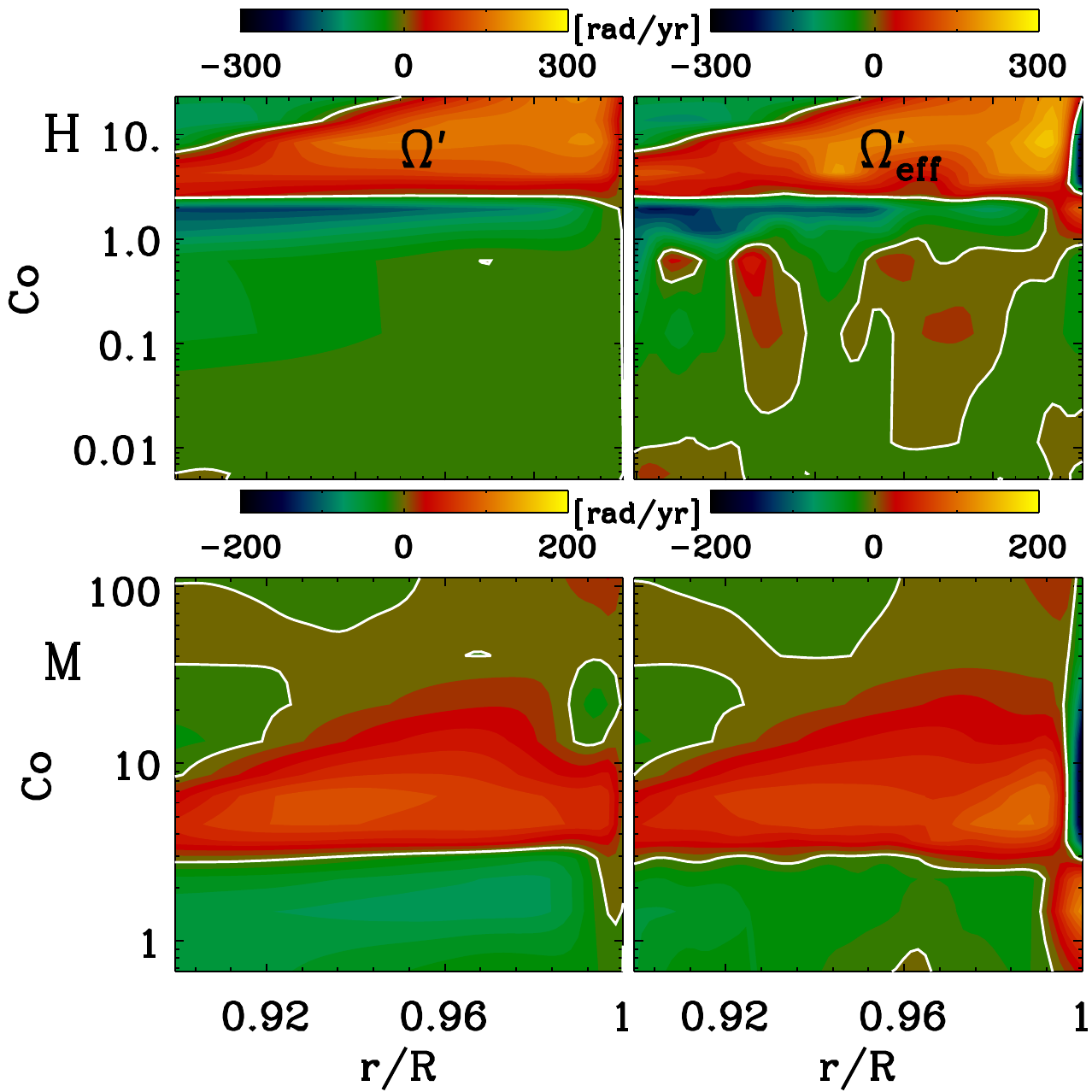}
\end{center}\caption[]{
The variation of radial shear of the mean flow $\Omega^\prime$ and
the effective flow $\Omega^\prime_{\rm eff}$ with rotation. We plot
$\Omega^\prime$ (first column) and $\Omega^\prime_{\rm eff}$ (second column) at
25$^\circ$ latitude as a function of $\Co$  for the HD runs (Set~H) in the top row and
for the MHD runs (Set~M) in the bottom row. The zero
values are indicated with a white line.
}\label{shear_eff}
\end{figure}

As a next step, we  further investigate the $\aalpha$ tensor by looking
at each of the diagonal components.
For this, we show in \Fig{alpha_ani} their meridional profiles and
the rms values.
For slow rotation, until $\Co=0.7$, the diagonal components have similar
strengths, but for larger rotation their behaviors diverge. Here,
$\alpharr$ shows a distribution 
with positive(negative) values in the upper part and
negative(positive) values in the bottom of the convection zone in the
northern(southern) hemisphere; it has the strongest values at
  low latitudes and near the surface. Here,
$\alpharr$ is the dominating 
component for moderate rotation
($\Co=1$ to 11), in particular in the 
HD runs.
For the highest rotation rates
(Runs~M10 to M30), we find a thin layer of opposite sign at the
surface.
However, in this regime $\alpharr$ becomes very weak.

Similarly to $\alpharr$,
$\alphatt$ has its strongest values 
for moderate rotation
($\Co=1$ to 11) with larger values in the 
HD runs, but it 
remains subdominant to the other two diagonal components at all rotation rates.
Interestingly,
$\alphatt$ is the only diagonal component, 
which changes sign as a function of rotation:
For $\Co<1.5,$ $\alphatt$ is dominantly negative(positive) in the 
northern(southern) hemisphere at low to mid
latitudes. For $\Co>1.5,$
it becomes positive (negative),
and finally approaches zero, being weaker than $\alpharr$ for the highest rotation rates.
This change was also reported by \cite{VKWKR19} by comparing their run
with $\Co=2.8$ to the run of \cite{WRTKKB17} with $\Co=8.3$.
At high latitudes, the distribution is similar as for $\alphaK$.

We find that $\alphapp$ follows the sign distribution of $\alphaK$ for values of
rotation, except for zero rotation. 
Its maximum values move from high 
to low latitudes as the rotations becomes stronger,
similarly to $\alpharr$.
An interesting finding is that, while $\alpharr$ dominates the moderately rotating runs
($\Co=1$ to 11), $\alphapp$ dominates the rapidly rotating ones.
For the highest rotation rates, the peak values are three to four times larger and
the rms values
are even five to eight times larger than for $\alpharr$ and 
$\alphatt$.

All in all, the $\aalpha$ tensor is highly anisotropic for all rotation rates 
above $\Co \approx 1$. The nature of the anisotropy changes from the
moderate dominance of $\alpharr$ to the strong dominance of $\alphapp$
for $\Co>10$. 
For rapidly rotating stars, spherical coordinates are not optimal and
the anisotropy of the $\aalpha$ tensor is even more easily seen if we
remap it to cylindrical coordinates ($\rho,\phi,z$).
Then, as shown in \Fig{alpcyl}, $\alpha_{zz}$ becomes close to zero.
This is in agreement with theoretical predictions \citep{R78,KR80} and
also roughly agrees with the axi- to nonaxisymmetric dynamo solution
transition found by \cite{VWKKOCLB17} (their limiting $\Co$ having
been around 3).
 However, the latter study did not report any quantitative change of the
 dynamo solutions at higher rotation rates, where according to our 
 results the anisotropies should play the most important role. 
 In any case, it seems that anisotropies in the $\alpha$ effect might play some
 role in the generation of the nonaxisymmetric modes.
This coexistence confirms the
mean-field calculation \citep[e.g.,][]{RWBMT90,ER07,P17}, where an
anisotropic $\aalpha$ tensor can generate nonaxisymmetric fields.

\subsection{Turbulent pumping}

\begin{figure*}[t!]
\begin{center}
\includegraphics[width=\textwidth]{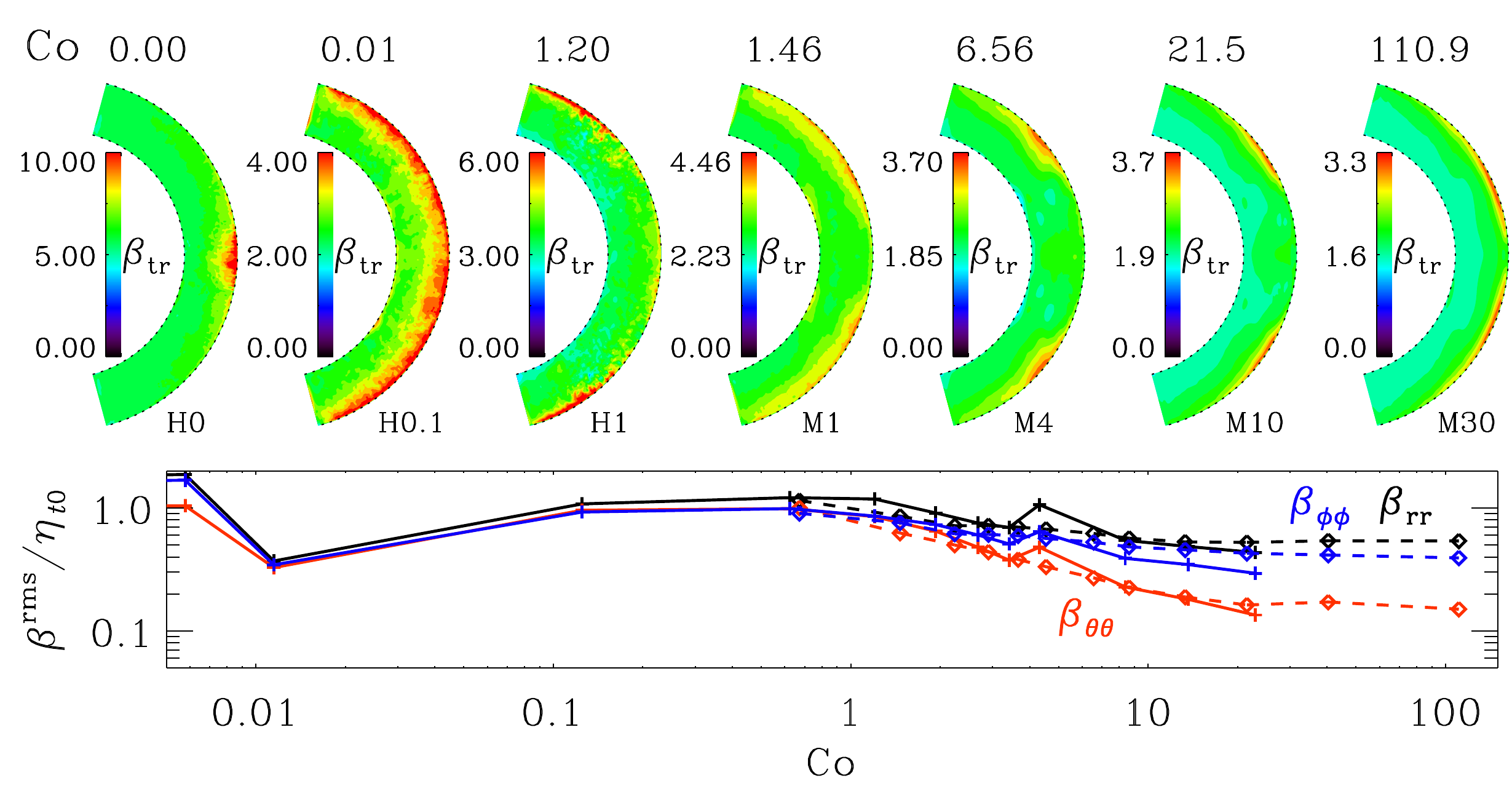}
\end{center}\caption[]{
Rotational dependency of the turbulent diffusion as the trace of $\bbeta$. We show
meridional profiles of $\beta_{\rm tr}$ (top row) and the rms values of
the diagonal component (bottom row), with
$\beta_{rr}$ (black lines), $\beta_{\theta\theta}$ (red), and
$\beta_{\phi\phi}$ (blue). 
As in \Fig{alpha_trace}, the values of the HD runs (Set~H) are shown
with a solid line and crosses, whereas the MHD runs (Set~M) are shown
with a dashed line with diamonds.
All values are normalized by $\etatz=\tau_{\rm tur} u^{\prime\,2}_{\rm
  rms}/3$.
The zero rotation run has been moved to $\Co=10^{-4}$ to be visible in the lower panel.
}\label{beta}
\end{figure*}

Now, we investigate the rotational influence of the turbulent pumping
vector $\ggamma$: see Fig.~\ref{gamma}. All of its components 
have a much weaker dependence on rotation than those of $\aalpha$.
The rms value of $\gamma_r$ increases
slightly with rotation up to 
$\Co$=2 and then decreases for higher
rotation. For all runs, we find 
an upward pumping near the surface,
agreeing with previous studies
\citep{WRTKKB17,VKWKR19}. 
However, some changes of the spatial profile can be distinguished as increasing
rotation: For slow rotation, we find the tendency for upward
pumping in the bulk of the convection zone and downward pumping near the bottom.
For $\Co$=6.6, $\gamma_r$ is  also pointing downward at high latitudes.
Furthermore, $\gamma_\theta$ also does not depend strongly on rotation.
 The HD runs exhibit some non-monotonic behavior in the form of an abrupt increase at 
 around $\Co$=6.6, but this bump is absent in the MHD runs. For all runs with rotation, the spatial distribution shows  equatorward pumping near the surface
of the upper part of the convection zone and poleward pumping near the
bottom.
$\gamma_\phi$ shows the strongest rotational dependency of all the turbulent pumping coefficients. 
For rotation rates up to $\Co=1.2$, $\gamma_\phi$  increases and for
higher $\Co$ it decreases, in particular in the MHD cases.
Again, the HD cases show non-monotonic behavior at the
same rotation rate as $\gamma_\theta$.
The spatial structure seems to be mostly independent of rotation. For
all runs with rotation, the pumping is prograde near the surface and at
the bottom of the convection zone, and weakly retrograde in the bulk of
the convection zone.
All values of $\ggamma$ are weaker than the turbulent velocity
$\urmsp$; for most cases they are only around 10 to 20\% of $\urmsp$.

To investigate how the turbulent pumping 
influences the
evolution of the magnetic field, we calculate the effective velocity,
$\meanUU^{\rm  eff}=\meanUU+\gamma$, 
which the magnetic field is sensitive to.
As shown in previous studies
\citep{WRTKKB17,VKWKR19}, $\ggamma$ can have a large impact on
$\meanUU^{\rm eff}$.
To this end,  \Fig{meri+diff_rot_eff} shows the effective meridional
and azimuthal flow (differential rotation) together with the original
meridional and azimuthal flow.
The azimuthal turbulent pumping is too weak to  significantly alter $\meanU_\phi^{\rm
  eff}$. However, we find that the anti-solar
differential rotation in the MHD runs is weaker due to
$\gamma_\phi$.
For the HD runs, $\meanU_\theta^{\rm eff}$ only changes for the
solar-like differential rotation. There, the multi-cellular structure
is altered to a noncellular structure.
For the MHD runs, the influence of $\gamma_\theta$ on
$\meanU_\theta^{\rm eff}$ is drastic. The equatorward flow in the
lower part of the convection zone completely vanishes and  even becomes
equatorward for some rotation rates. Moreover, the other flow structures
are significantly altered.

Even though we do not find strong changes in $\meanU_\phi^{\rm
  eff}$ due to $\gamma_\phi$, the radial shear can nevertheless
  change \citep{WRTKKB17}.
To check for this possible effect, we plot in \Fig{shear_eff}  the radial shear 
defined as
$\Omega^\prime=r\sin\theta\,\partial\Omega/\partial r$ and $\Omega^\prime_{\rm
  eff}=r\sin\theta\,\partial\Omega^{\rm eff}/\partial r$ with $\Omega^{\rm
  eff}=\meanU_\phi^{\rm eff}/r\sin\theta + \Omega_0$.
For weak rotation, we find that the 
dominantly negative shear changes
to positive values in the bulk of the convection zone for slow to moderate rotation rates.
More dramatic are the changes near the surface. For nearly all runs,
an additional layer of shear of opposite sign 
to the overall shear in the bulk
is generated near the very surface.
For slow rotation rates, a positive shear region appears, while for high rotation,
a negative one appears.
Hence, our results support the conclusion of \cite{WRTKKB17} and \cite{VKWKR19}, namely that
the turbulent pumping plays an important role for the 
evolution of the magnetic field.

\subsection{Turbulent diffusion and the
  R\"adler effect}

\begin{figure}[t!]
\begin{center}
\includegraphics[width=\columnwidth]{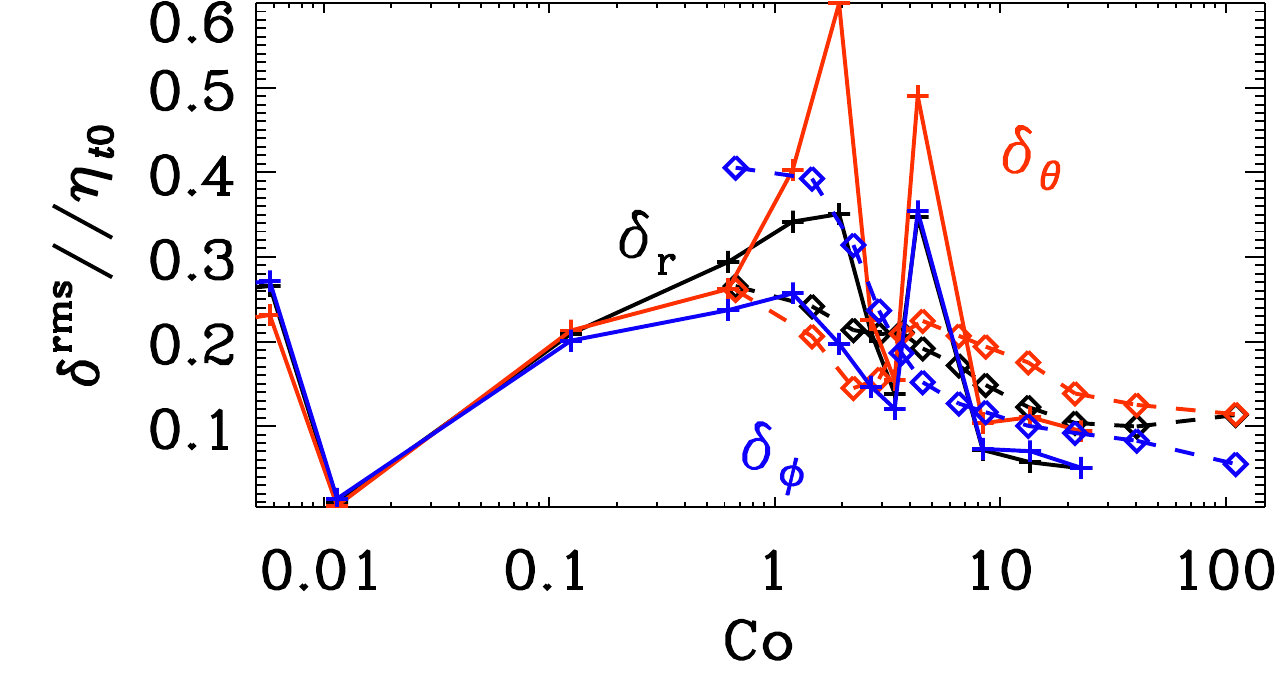}
\end{center}\caption[]{
Rotational dependency of the R\"adler effect. We show the rms values of
the component of $\ddelta$ with
$\delta_{r}$ (black lines), $\delta_{\theta}$ (red), and
$\delta_{\phi}$ (blue). 
As in \Fig{alpha_trace}, the values of the HD runs (Set~H) are shown
with a solid line and crosses, whereas the MHD runs (Set~M) are shown
with a dashed line with diamonds.
All values are normalized by $\etatz=\tau_{\rm tur} u^{\prime\,2}_{\rm
  rms}/3$.
The zero rotation run has been moved to $\Co=10^{-4}$ to be visible.
}\label{delta}
\end{figure}

To investigate the rotational dependency of turbulent diffusion, we
limit ourselves to the diagonal components and the trace of $\bbeta$, $\beta_{\rm
  tr}=\beta_{rr}+\beta_{\theta\theta}+\beta_{\phi\phi}$.
As shown in \Fig{beta}, $\beta_{\rm tr}$ decreases with rotation and
we find that it is often two times stronger
near the surface
than in the lower part of the convection zone. The trace is always
larger than zero, with the exception of 
some values at the radial and latitudinal boundaries,
which might be artifacts.
The rms values of the diagonal components 
remain roughly independent of rotation until $\Co=7$.
In this regime, all components have strengths
close to $\etatz$.
For higher rotation, all three
diagonal components decrease with rotation up to $\Co=22$, 
the decrease being most pronounced for the
$\theta\theta$ component whose 
values become roughly
four times smaller than
$\beta_{rr}$ and three times smaller
than $\beta_{\phi\phi}$.
For even
higher rotation, the $\beta$ components remain roughly constant.
As for the $\aalpha$ tensor, we also map the $\bbeta$ tensor to
  cylindrical coordinates and plot its diagonal components in
  \Fig{betacyl}. The differences between the presentations 
  in the two coordinate  systems are not large. 
Now $\beta_{zz}$ is the weakest of all three diagonal
components, whereas $\beta_{\rho\rho}$ and $\beta_{\phi\phi}$ have 
similar values. This is in agreement with the theoretical prediction,
according to which $\beta_{zz}$ should be most highly quenched for rapid
rotation \citep{KR80}.

Now, we turn to the R\"adler effect, expressed by $\ddelta$.
As shown in \Fig{delta}, all components of $\ddelta$ 
increase for slow and decrease for fast rotation.
For HD runs, $\delta_\theta$ becomes strong for the runs with a
strong energy in differential rotation  $E_{\rm kin}^{\rm dif}$: in the anti-solar
regime, Run~H1.5, and in the solar-like regime, Run~H3.
 Furthermore, $\delta_r$ is also strong in the former run 
and $\delta_\phi$ in the latter.
For $\Co>4,$ $\delta_\theta$ is dominating the other
components in both HD and MHD runs.
The values of all delta components are of the order of 10-30\% of $\etatz$.

\begin{figure}[t!]
\begin{center}
\includegraphics[width=\columnwidth]{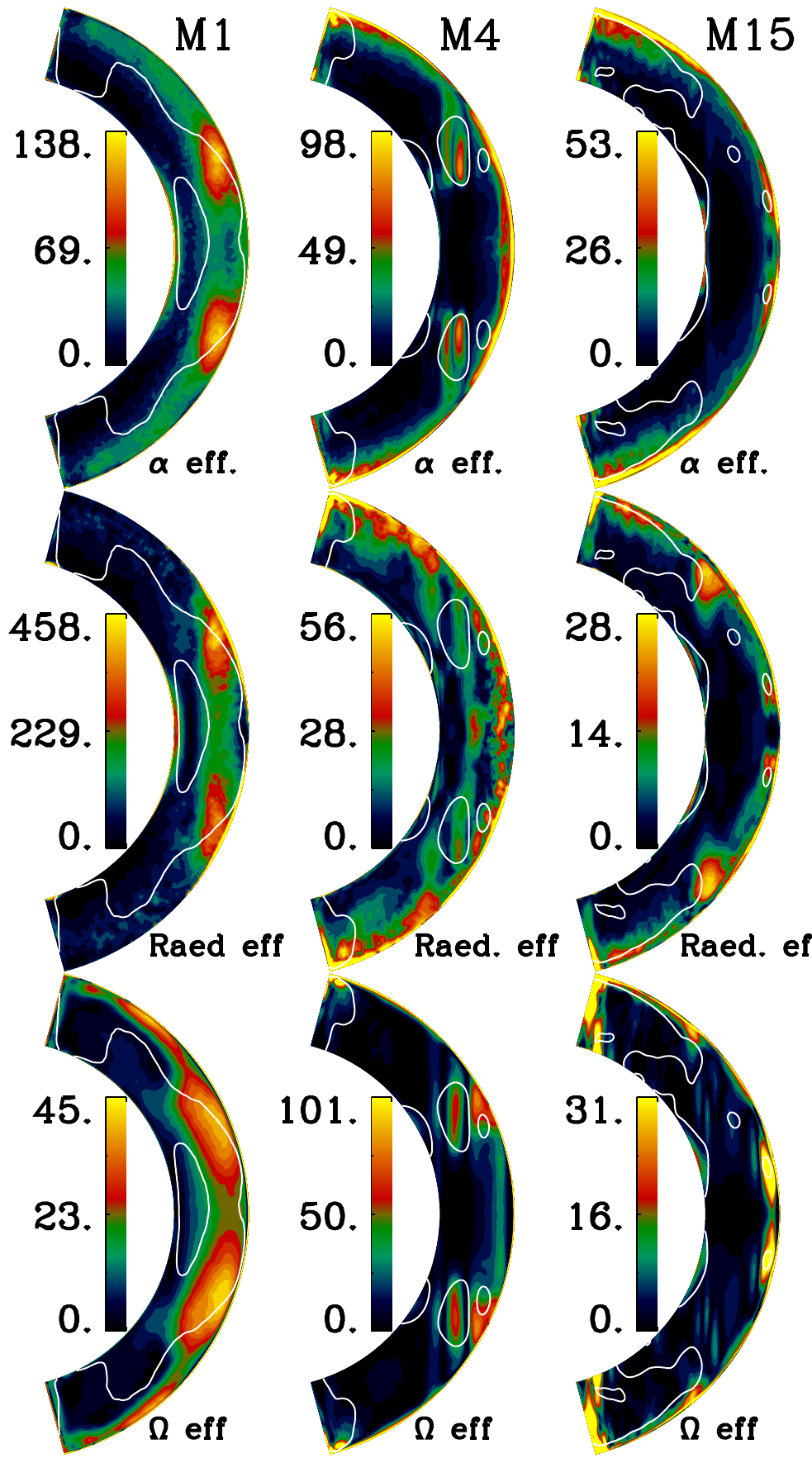}
\end{center}\caption[]{
Dynamo mechanism generating $\meanB_\phi$ for Runs~M1, M4, and M15.
We show the $\alpha$ effect (top row), the R\"adler/$\delta$
effect (middle), and the $\Omega$ effect (bottom), where we overplot as
white contours the rms values of $\meanB_\phi$ above half of the
maximum, indicating the magnetic field region used in the calculation
of the dynamo mechanism shown in \Fig{dyndrivB}. All values are given in
kG/yr.
}\label{dyndrivA}
\end{figure}

\begin{figure}[t!]
\begin{center}
\includegraphics[width=\columnwidth]{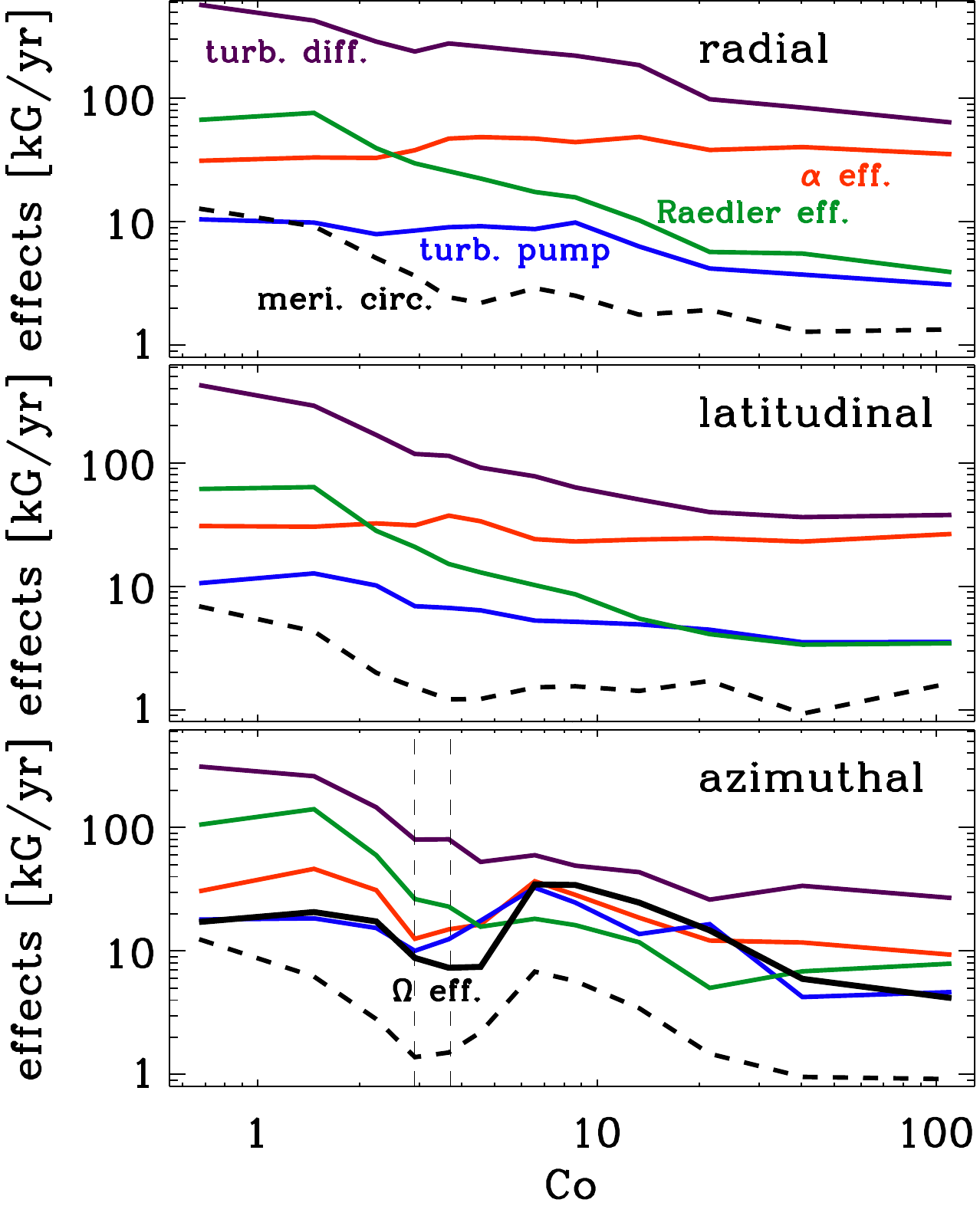}
\end{center}\caption[]{
Rotational dependence of the main dynamo mechanism driving the radial
field (top panel), the latitudinal field (middle), and the
azimuthal field (bottom) with the $\alpha$ effect (red), the turbulent
diffusion (purple), the turbulent pumping (blue), the meridional
circulation (black dashed), the R\"adler effect (green), and the
$\Omega$ effect (black solid). The vertical dashed lines indicate the
transition from the anti-solar to solar-like differential rotation. See \Sec{sec:dyndriv} for the
calculation details.
}\label{dyndrivB}
\end{figure}

\subsection{Rotational dependency of the dynamo mechanism}
\label{sec:dyndriv}

To test which dynamo effects are responsible for
the generation and evolution of the magnetic field, we 
individually monitor the following terms in the induction equation: 
\begin{eqnarray}
&\nab\times\aalpha\cdot\meanBB&\quad\text{$\alpha$ effect},\\
&\nab\times\ggamma\times\meanBB&\quad\text{turbulent pumping},\\
&\nab\times\bbeta\cdot\nab\times\meanBB&\quad\text{turbulent diffusion},\\
&\nab\times\ddelta\times\nab\times\meanBB&\quad\text{R\"adler effect},\\
&\nab\times(\meanuu_{\rm dif}\times\meanBB)&\quad\text{$\Omega$ effect},\\
&\nab\times(\meanuu_{\rm mer}\times\meanBB)&\quad\text{meridional
                                                  circulation},
\end{eqnarray}
where $\meanuu_{\rm dif}=(0,0,\mean{u_\phi})$ and $\meanuu_{\rm
  mer}=(\mean{u_r}, \mean{u_\theta}, 0)$. We then take the temporal rms
$\sqrt{\bbra{\cdot^2}_t}$ of each of these terms.
In \Fig{dyndrivA}, we show an example of the field generators for $\meanB_\phi$
from the $\alpha$, R\"adler, and $\Omega$ effects from three different runs. 
We see for Run~M1
that the strongest effect is the R\"adler effect, which 
acts on the same location as the $\alpha$ effect. The mean azimuthal field,
the strongest field regions being indicated with white contour lines, actually 
concentrates closer to the equator and nearer to the bottom of the convection zone 
than the distribution of the field generators.
This can be explained by the meridional pumping which is equatorward at
these locations, added with downwards-directed $\gamma_r$ in the 
bottom parts of the convection zone.
For M~4 and M10 with $\Co$=6.6 and 22, 
oscillatory magnetic fields with clear equatorward migration are excited,
as can be seen from \Fig{but}.
From \Fig{dyndrivA} we see that the $\Omega$ 
and the $\alpha$ effects are strong in the areas where the azimuthal field is also
large.
For even higher rotation in Run~M15,
the field generation near the surface near
the equator is mostly due to the $\alpha$ effect with a contribution
from the $\delta$ and $\Omega$ effects. For the high-latitude fields,
$\alpha$ and R\"adler effects have stronger contributions.
The magnetic field evolution shows an
irregular solution with an
indication of poleward migrating field which might fit to this kind of dynamo.
This result also shows that even if the $\aalpha$ tensor is highly
anisotropic with very low values for $\alpharr$ and $\alphatt$, the
$\alpha$ effect generating the azimuthal field is still strong and can
sustain a strong large-scale dynamo. 

To further refine our analysis, we adopt the approach by \cite{W18} to measure
the dynamo effects only from the locations where the magnetic field component
that they are acting on    is  larger than the half maximum of its rms value
(indicated by white contour lines in \Fig{dyndrivA}). 
We average each dynamo effect over these 
locations and plot all of them as a
function of rotation in \Fig{dyndrivB}.

From Fig. 13 we
find that the dynamo effects show a strong rotational dependence.
The effect of turbulent diffusion
decreases for all runs for increasing rotation, even though the
diagonal components of $\bbeta$
become constant.
This is most likely
due to the decrease of turbulent intensity with rotation, which 
is used as a normalization in \Fig{beta}.
Except for slow rotation ($\Co$=0.7 to 1.5), the effect of
the turbulent pumping is significantly stronger than the effect on the
meridional circulation. Both effects
decrease with increasing rotation.

For the other effects, we find three distinct regimes. 
The first regime is for slow rotation, where the differential rotation
is anti-solar. There, the R\"adler and the $\alpha$ effects are
dominating even over the $\Omega$ effect for azimuthal field generation.
This would rule out an $\alpha$$\Omega$ effect and suggest an
$\alpha^2$-type dynamo with a strong $\delta$ contribution.
From our method of using averaged rms values without signs, 
we cannot conclude whether the R\"adler effect contributes to the magnetic
field enhancement or the diffusion of the field, but nevertheless our
analysis indicates a strong role of this effect in the evolution of the magnetic field.
This regime is consistent with the finding of \cite{VKWKR19},
where the authors find indications for an
$\alpha^2$-type model with a strong $\delta$ contribution  
in their run with $\Co$=2.8.
However, these latter authors find a cyclic magnetic field solution in contrast to our
stationary or irregular ones, as shown in \Fig{but}.
For the runs in the transitionary phase of the differential rotation
(Runs~M2 to M3, with $\Co$=2.9 to 4.5), the $\Omega$ effect is even
weaker than for slow rotation, and the dynamo is dominated by an
$\alpha^2$ contribution with a much weaker $\delta$. However, we see
clearer indication of cycles, which are nevertheless not yet very pronounced; see \Fig{but}.

The second regime is one where the runs exhibit solar-like
differential rotation profiles. 
There, the $\Omega$ effect is comparable to or even dominates 
the $\alpha$ effect in generating the azimuthal field.
Runs in this regime (M4 to M10 with $\Co$=6.6
 to 22) show clear
equatorward migration of the oscillating magnetic fields.
Previous analyses of similar runs have revealed that these dynamo waves
can be explained by the Parker-Yoshimura rule \citep{WRTKKB17,W18}
for $\alpha\Omega$ dynamos.
This is consistent with our finding of
the $\Omega$ effect being large in \Fig{dyndrivB}. The strong
contribution that we find for $\alpha$ in generating the azimuthal field can still
contribute to the magnetic field evolution, but it does not
necessarily influence the period and propagation direction of the
dynamo wave, 
as was the case also in the studies of 
\cite{WRTKKB17} and \cite{W18}.

The third regime is observed
for runs with the highest rotation rates
(M15 \& M30, with $\Co$=40 \& 111). The R\"adler effect has 
decreased with rotation for
the poloidal components of the field, while the $\alpha$ effect
has remained roughly
constant. For the azimuthal component, the $\Omega$ effect 
drops to
much lower values, as already indicated 
by the decrease of the overall energy in the differential rotation in \Fig{energies}.
The R\"adler effect for this component is comparable
to the $\alpha$ effect. This indicates an
$\alpha^2$ dynamo with a weak
$\delta$ contribution for the azimuthal magnetic field.

With the rotational dependence of the dynamo effects, we can now
understand why the  ratio of magnetic to kinetic energy increases even though the shear and
normalized $\aalpha$ tensor remain constant for moderate to high
rotation. 
This is because simultaneously turbulent diffusion decreases
as a function of rotation, enabling a more efficient dynamo action.
This is in very good agreement with the results of \cite{KKB09}, who
find similar behavior from turbulent convection in Cartesian domains
where rotational influence was varied. 

\section{Conclusions}

We performed a comprehensive study of how turbulent transport
coefficients measured from global convective dynamo simulations
depend on rotation in solar-like stars.
For this, we varied 
the rotational influence of convection in
terms of Coriolis number from $\Co$=0 to $\Co$=110.
We found that 
the normalized trace of $\aalpha$ only 
increases up to
$\Co=4$, with 
an approximate power law of $\Co^{0.5}$, and then levels off.
The trace of $\aalpha$ 
shows a very similar spatial profile in comparison to an
expression of $\alpha$ based on
the kinetic helicity, $\alphaK$.
However, this quantity does not level off, but continues its growth even
in the rapid rotation regime. However, if we take into account that the length
scales of convection become reduced with increasing rotational influence,
the effect of which could result in a decrease of the correlation time with a power
law of $\Co^{-0.5}$ according to theoretical considerations
\citep{Ch61} and recent numerical results \citep{FH16,VWKKOCLB17},
then $\alphaK$ also levels off. 
The magnetic correction to the $\alpha$ effect,
expressed in terms of $\alphaM$ based on the current
helicity, 
becomes anomalously strong in the cases of rapid rotation, even
exceeding the value of $\alphaK$  locally. Therefore, it is not justified to 
consider $\alphaM$ as a perturbation,
as in the original analysis of \cite{PFL76}.
However, we treated the turbulent correlation times for the flow
and magnetic fields equally in our analysis, which might explain
the discrepancy. This issue  requires further investigation, but
even at this stage it is clear that some caution is needed when deriving the turbulent
transport coefficients from convection simulations using these proxies.

We further find that the $\aalpha$ tensor becomes highly 
anisotropic
for $\Co$>1, as expected from theoretical predictions \citep{KR80}.
In the moderate rotation regime, $\alpharr$ dominates over the
other diagonal components.
The nature of the anisotropy changes for $\Co$>10, when 
$\alpharr$ and $\alphatt$ become strongly reduced, while $\alphapp$
strongly increases.
Anisotropies in the $\aalpha$ tensor are one of the candidates leading
to nonaxisymmetric large-scale dynamo solutions \citep[e.g.,][]{RWBMT90,ER07,P17}. 
However, the transition to nonaxisymmetry was seen at somewhat elevated Coriolis numbers of roughly three
in the study of \cite{VWKKOCLB17}, where similar runs to those here  were reported, but
with the full longitudinal extent and therefore capable of naturally exciting
nonaxisymmetric large-scale modes. The difficulty in analyzing 
such runs with the test-field method arises from the fact that the axisymmetric
averages are not suitable for nonaxisymmetric dynamo solutions. Confirmation
of the importance of $\alpha$ effect anisotropies in the excitation of nonaxisymmetric
modes must therefore await further development of the test-field method and/or
appropriate 3D mean-field modeling taking into account the dynamo effects measured here.

The turbulent pumping components
do not strongly depend on rotation. 
In all runs, we measure 
upward pumping near the surface, in contrast to what 
is needed for
the surface-flux-transport 
models to agree with solar
observations \citep[e.g.,][]{CSJ12}. The latitudinal pumping is mostly
equatorward in the upper part of the convection zone and poleward in
the lower part, and therefore it could be able to advect a dynamo wave
equatorward if it overcame diffusion.
However, this is not seen in 
any of our models. The 
presence of the latitudinal pumping completely
alters the effective meridional circulation, 
which has strong implications for flux-transport dynamo models
that rely on certain types of meridional circulation profiles and do
not fully consider all the turbulent effects.
The azimuthal pumping leads to a sharp sign change for the effective
shear near the surface for all runs.

We find that the normalized turbulent diffusion 
decreases slightly with rotation before it levels off at around
$\Co$>10 and becomes weakly anisotropic.
$\ddelta$, describing
the R\"adler effect, is the strongest for moderate
rotation, where differential rotation is also the strongest.

Analyzing the dynamo effects as a function of rotation reveals three
distinct
regimes. For slow rotation, we find strong $\alpha$ and R\"adler
effects together with anti-solar differential rotation, consistent
with the work of \cite{VKWKR19}. 
For moderate rotation, where differential rotation is solar-like 
and the magnetic field develops equatorward migration with clearly
defined cycles, we find a strong contribution from both $\alpha$ and
$\Omega$ effects for the generation of the toroidal component
and from the
$\alpha$ effect for the generation of the poloidal component, while the
other effects remain subdominant. This is consistent 
with an $\alpha\Omega$ or $\alpha^2\Omega$ dynamo, in agreement with previous
studies \citep{WRTKKB17,W18}. 
For high rotation, the $\alpha$ effect contributions remain strong, while the $\Omega$ effect
is significantly reduced.
Therefore, our interpretation is that a dynamo of
$\alpha^2$ type, with some influence of the R\"adler effect for
azimuthal field, is operating in this regime.
The dynamo efficiency, defined as the ratio of
magnetic energy to kinetic energy, increases linearly with $\Co$, in
agreement with previous studies \citep{VWKKOCLB17,ABT19}.
This can be explained by the $\alpha$ effect being almost independent
of rotation, whereas the effect of turbulent diffusion decreases with
rotation in agreement with the Cartesian models of \citep{KKB09}.

\begin{acknowledgements}
We thank the anonymous referee and our colleague Petri J. K\"apyl\"a for
constructive comments on the manuscript.
The simulations have been carried out on supercomputers at
GWDG, on the Max Planck supercomputer at RZG in Garching and in the facilities hosted by the CSC---IT
Center for Science in Espoo, Finland, which are financed by the
Finnish ministry of education. 
J.W.\ acknowledges funding by the Max-Planck/Princeton Center for
Plasma Physics and from the People Programme (Marie Curie
Actions) of the European Union's Seventh Framework Programme
(FP7/2007-2013) under REA grant agreement No.\ 623609.
This project has received funding from the European Research Council
(ERC) under the European Union's Horizon 2020 research and innovation
programme (grant agreement n:o 818665 ``UniSDyn''), and has been
supported from the Academy of Finland Centre of Excellence ReSoLVE
(project number 307411).

\end{acknowledgements}

\bibliographystyle{aa}
\bibliography{paper}

\appendix

\section{Rotation profiles and butterfly diagrams for all runs}

\begin{figure}[t!]
\begin{center}
\includegraphics[width=0.75\columnwidth]{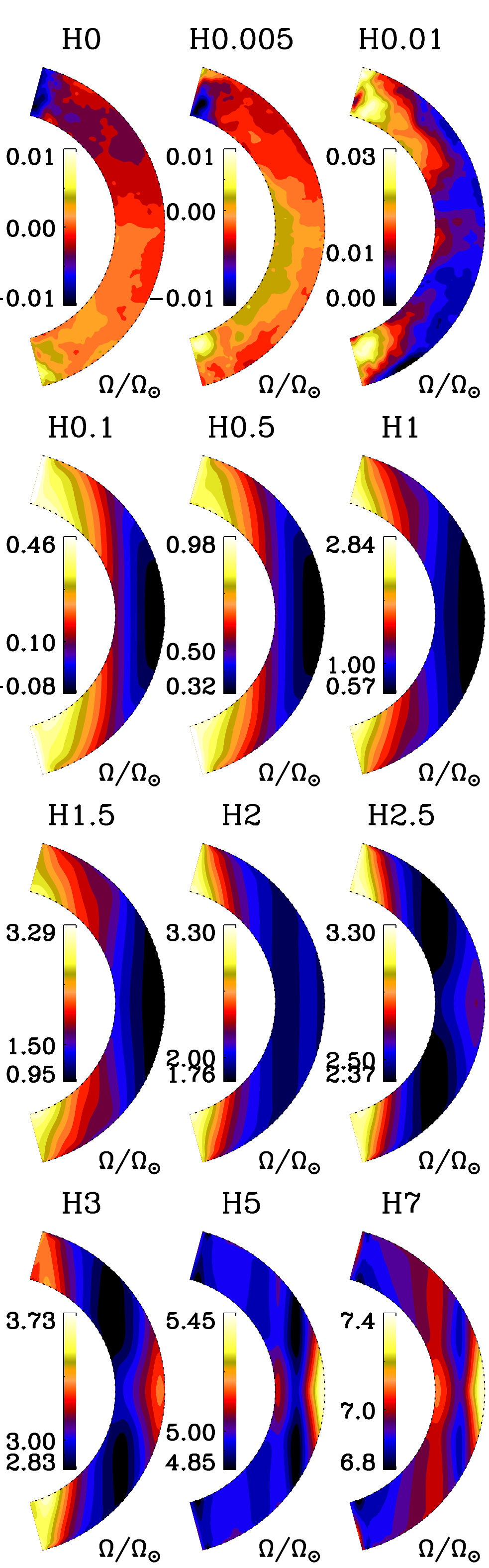}
\end{center}\caption[]{
Normalized local rotation profile $\Omega/\Omega_\odot$ with
$\Omega=\Omega_0+\meanuu/r\sin\theta$ for all HD runs (Set~H), except run H~10. The value $\Omega$ has been calculated as a time average over the
saturated state.
}\label{diffrot_HD}
\end{figure}

\begin{figure}[t!]
\begin{center}
\includegraphics[width=0.75\columnwidth]{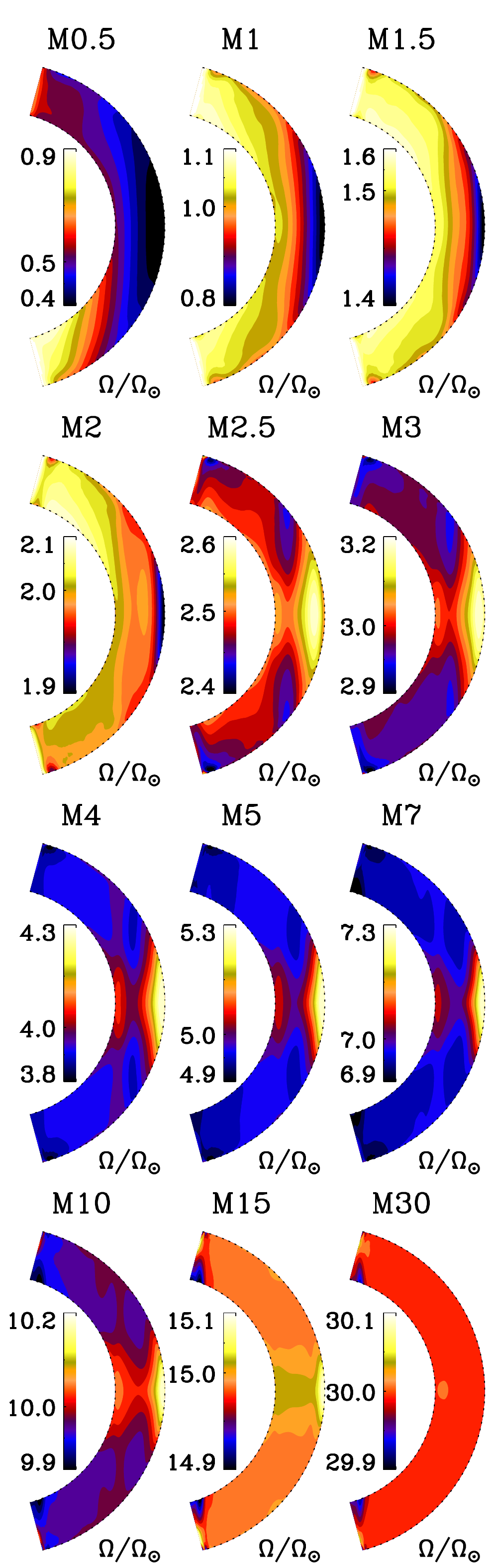}
\end{center}\caption[]{
Normalized local rotation profile $\Omega/\Omega_\odot$ with
$\Omega=\Omega_0+\meanuu/r\sin\theta$ for all MHD runs (Set~M). The
value $\Omega$ has been calculated as a time average over the saturated state.
}\label{diffrot_MHD}
\end{figure}

\begin{figure}[t!]
\begin{center}
\includegraphics[width=\columnwidth]{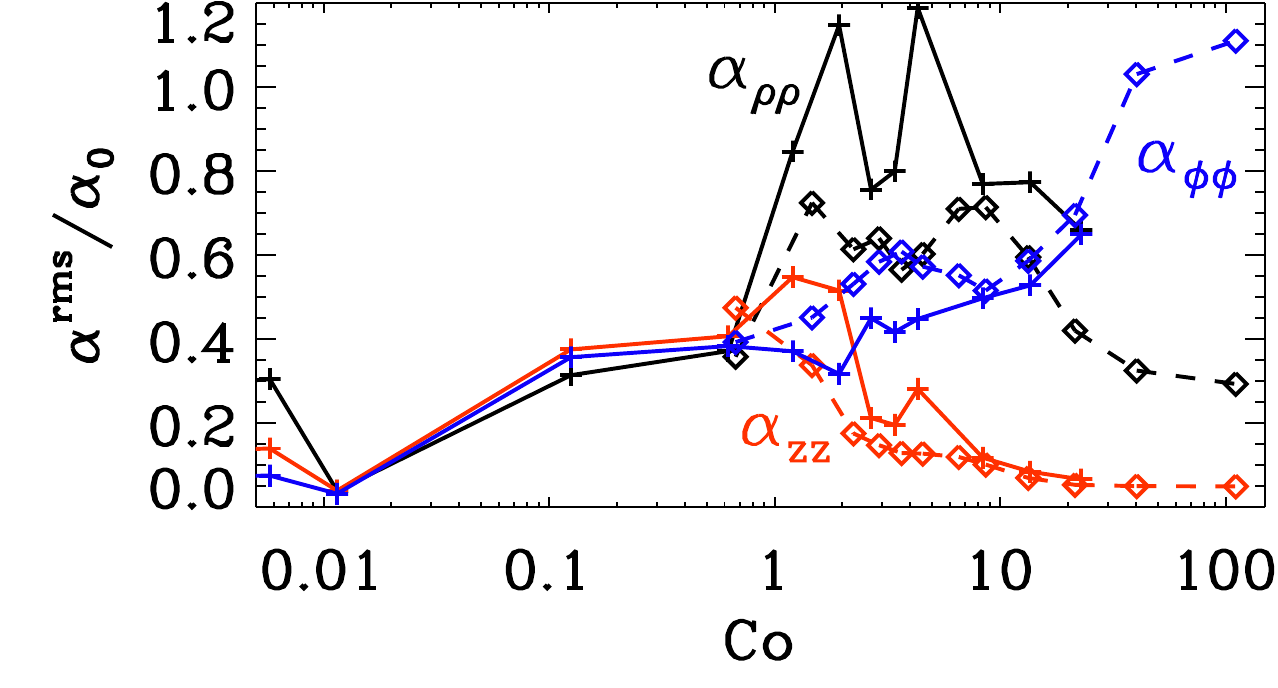}
\end{center}\caption[]{
Same plot as the bottom panel of \Fig{alpha_ani} but for the diagonal
components of $\aalpha$ in cylindrical coordinates ($\rho,\phi,z$) with
$\alpha_{\rho\rho}$ (black lines), $\alpha_{zz}$ (red) and $\alphapp$ (blue).
}\label{alpcyl}
\end{figure}

For completeness,  in this Appendix we present the rotation profiles
(Fig.~\ref{diffrot_HD} for HD runs (Set~H) and Fig.~\ref{diffrot_MHD}  for MHD
runs (Set~M)) and butterfly diagrams (Fig.~\ref{but}) from all the runs
analyzed in the main part of the paper. 
In \Figs{alpcyl}{betacyl}, 
we additionally show the diagonal components of the $\aalpha$ tensor and
$\bbeta$ tensor, respectively, in cylindrical coordinates to aid
comparison with theoretical studies.

\begin{figure}[t!]
\begin{center}
\includegraphics[width=\columnwidth]{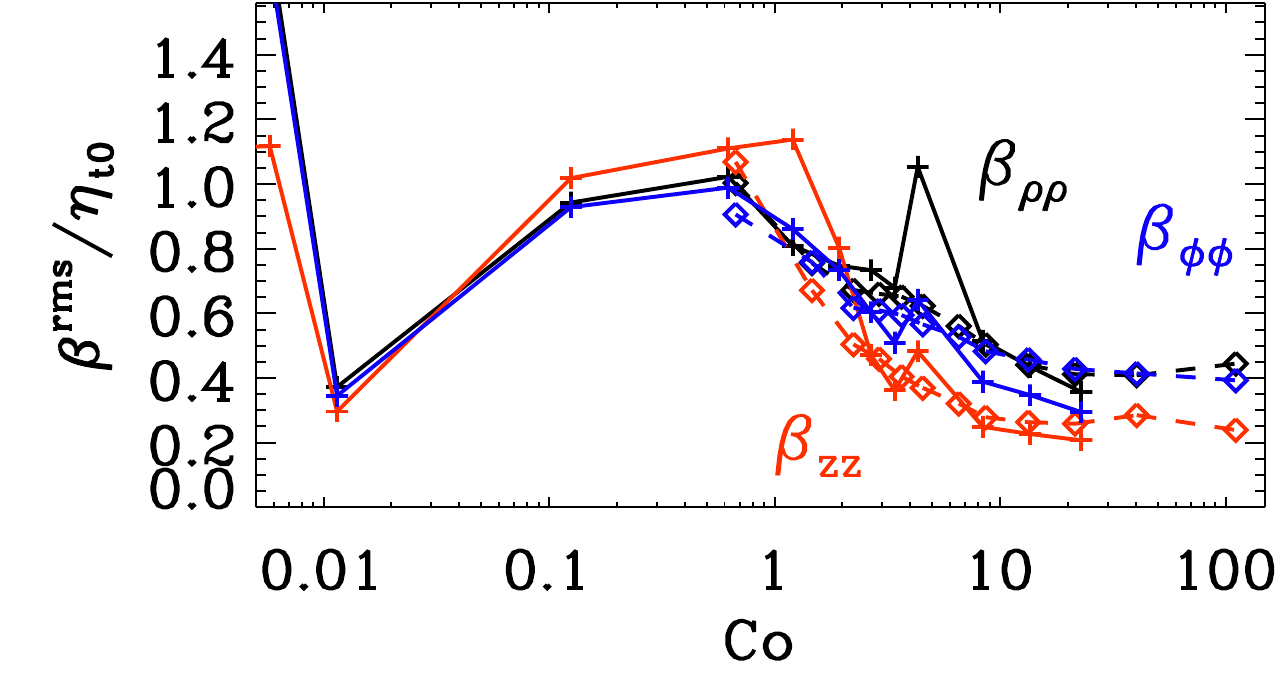}
\end{center}\caption[]{
Same plot as the bottom panel of \Fig{beta} but for the diagonal
components of $\bbeta$ in cylindrical coordinates ($\rho,\phi,z$) with
$\beta_{\rho\rho}$ (black lines), $\beta_{zz}$ (red) and
$\beta_{\phi\phi}$ (blue).
}\label{betacyl}
\end{figure}

\begin{figure*}[t!]
\begin{center}
\includegraphics[width=\textwidth]{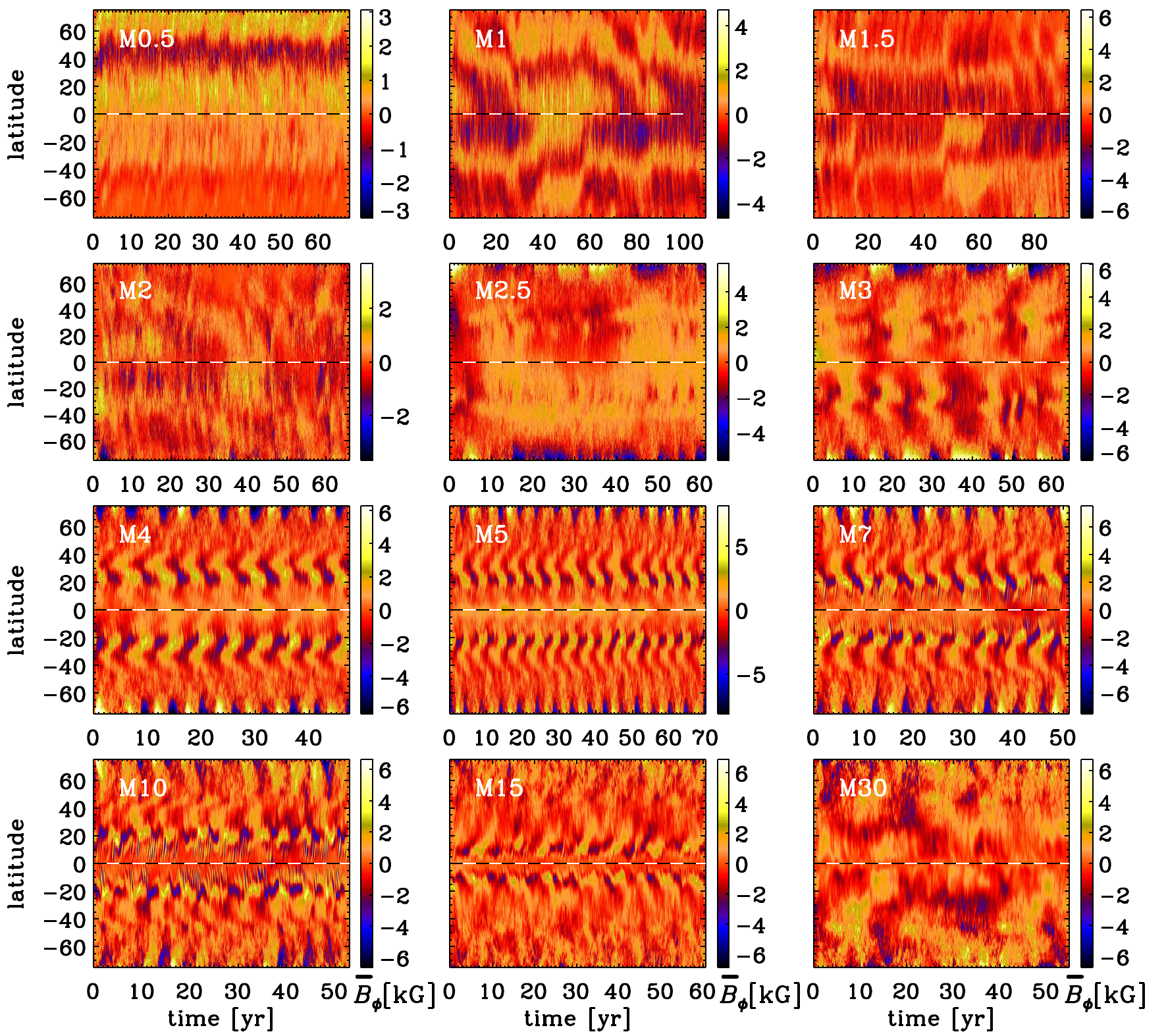}
\end{center}\caption[]{
Mean azimuthal magnetic field $\meanB_\phi$ as a function of time in
years and latitude near the surface ($r=0.98R$) for all MHD runs (Set~M). The time interval shows the full
duration of the saturated state for all runs. The black and white dashed horizontal line indicates the equator.
}\label{but}
\end{figure*}

\end{document}